\documentclass[journal=jctcce,manuscript=article]{achemso}

\makeatletter
\renewenvironment{figure}[1][]{\@float{figure}[#1]}{\end@float}
\renewenvironment{figure*}[1][]{\@dblfloat{figure}[#1]}{\end@dblfloat}
\makeatother
\usepackage[version=3]{mhchem}
\usepackage{amsmath}
\usepackage{physics}
\usepackage{graphicx}

\usepackage{pdfpages} % for si.pdf inclusion

\newcommand{\cre}[1]{a_{#1}^{\dag}}
\newcommand{\ani}[1]{a_{#1}^{}}

\title{Can phaseless auxiliary-field quantum Monte Carlo with broken symmetry trials describe iron-sulfur clusters?}

\author{Eirik F.~Kjønstad}
\affiliation{Division of Chemistry and Chemical Engineering, California Institute of Technology, Pasadena, California 91125, USA}

\author{Huanchen Zhai}
\affiliation{Initiative for Computational Catalysis, Flatiron Institute, 160 Fifth Avenue, New York 10010, New York, USA}

\author{James Shee}
\affiliation{Department of Chemistry, Department of Physics and Astronomy, Rice University, Houston, Texas 77005-1892, United States}

\author{Sandeep Sharma}
\affiliation{Division of Chemistry and Chemical Engineering, California Institute of Technology, Pasadena, California 91125, USA}

\author{Garnet Kin-Lic Chan}
\email{gkc1000@gmail.com}
\affiliation{Division of Chemistry and Chemical Engineering, California Institute of Technology, Pasadena, California 91125, USA}

\begin{document}

\begin{abstract}
Phaseless auxiliary-field quantum Monte Carlo (AFQMC) has in several cases been found to perform well on strongly correlated systems, including compounds containing transition-metal elements. Here, we benchmark the method for three iron-sulfur clusters ([2Fe-2S], [4Fe-4S], and the FeMo cofactor) using a hierarchy of systematically improved trial states derived from coupled cluster (CC) theory, including up to quadruple excitations relative to the Hartree-Fock (HF) reference, as well as multi-Slater trial states derived from unrestricted density matrix renormalization group wavefunctions.
Our results reveal for these systems that, as the symmetry-broken trial is improved beyond the mean-field level, the phaseless AFQMC energy can become less accurate, and in some cases even less accurate than the underlying trial projected energy itself, displaying an inverted energy pattern that is only corrected once the trial fidelity is sufficiently high.
For [2Fe-2S], we show that a decreased accuracy in the AFQMC energy can coincide with a simultaneous improvement in the trial state and the walker ensemble, as measured by their fidelity relative to the exact ground state, a behavior which is allowed by the non-variational nature of the phaseless AFQMC energy estimator.
We further find that this is not solely due to the use of spin-unrestricted trial states, as the inversion persists in [2Fe-2S] when we explicitly break the symmetry of the Hamiltonian by applying a fictitious local spin-Zeeman field.
Instead, we find that the energy inversion is related to the choice of measurement trial in the estimator, where using a high-order CC trial state for measurements may introduce errors that are suppressed when the measurement wave function is restricted to lower excitation subspaces.
In particular, measuring the energy with the mean-field reference while guiding the walkers with a CC trial removes the inverted pattern and improves the overall accuracy across the iron-sulfur clusters, with a possible exception for [4Fe-4S] where energies with a CCSDT guide may be less accurate than those guided by CCSD for some broken-symmetry families.
Taken together, our findings suggest that the relatively accurate energies obtained with an HF trial state in these systems
arise from favorable error cancellation, warranting significant caution about the reliability of phaseless AFQMC with such trials
for strongly correlated transition-metal systems of this kind.
\end{abstract}

%%%MAIN TEXT%%%%

\section{\label{sec:intro}Introduction}
Iron-sulfur (Fe-S) clusters are ubiquitous inorganic cofactors in biological redox chemistry. They participate in various processes, including electron transfer, iron/sulfur storage, gene regulation, and enzymatic catalysis.\cite{johnson2005} The common structural motifs, [2Fe-2S], [3Fe-4S], and [4Fe-4S], support diverse and complex  reactivity, while a prominent example of a larger, eight metal, Fe-S cluster is the iron-molybdenum (FeMo) cofactor of the nitrogenase enzyme, which catalyzes the conversion of dinitrogen (\ce{N2}) to ammonia (\ce{NH3}) under ambient conditions.\cite{einsle2002nitrogenase, spatzal2011evidence}

From an electronic-structure perspective, Fe-S clusters appear challenging because they feature many near-degenerate Fe $3d$-derived orbitals and multiple low-lying spin manifolds arising from open-shell Fe centers and competing spin-coupling and orbital occupancy patterns.\cite{sharma2014low} The importance of electron correlation in the low-energy spectrum has made achieving chemical accuracy ($\sim$1 kcal/mol) in total and relative energetics particularly challenging.
As a result, Fe-S clusters have become canonical stress tests for classical multireference methodology and quantum-computing approaches, where resource estimation work frequently highlights the nitrogenase FeMo cofactor as a flagship target.\cite{reiher2017elucidating,lee2021even, lee2023evaluating,low2025fast} Despite the challenging nature of this system, recent work by some of the authors
has shown that, for an active space model\cite{li2019electronic} capturing the dominant physics, the ground-state energy can be estimated classically to within chemical accuracy using a combination of high-order coupled cluster and density matrix renormalization group (DMRG) calculations, albeit at substantial computational cost.\cite{zhai2026classicalsolutionfemocofactormodel}

The development of more efficient classical methods for strongly correlated transition-metal clusters like the FeMo cofactor is therefore an area of continued interest.
Among established electronic-structure methods, phaseless auxiliary-field quantum Monte Carlo (ph-AFQMC) using single determinant trials has been argued to perform better in correlated regimes than single-reference non-symmetry broken CC approaches.
Recent benchmarks and perspectives have therefore proposed phaseless AFQMC as a practical approach to obtain benchmark-quality data for challenging main-group and transition-metal systems,
potentially overcoming known shortcomings and the cost of exact CCSD(T)\cite{raghavachari1989fifth,shee2019achieving, shee2023potentially,motta2018ab} in these regimes.
 However, as an emerging method, the systematic accuracy of phaseless AFQMC has not yet been tested to the same extent across large and diverse benchmark sets, when compared with other established methods like CCSD(T).\cite{lee2022twenty}

\begin{figure}[!htbp]
    \includegraphics[width=\linewidth]{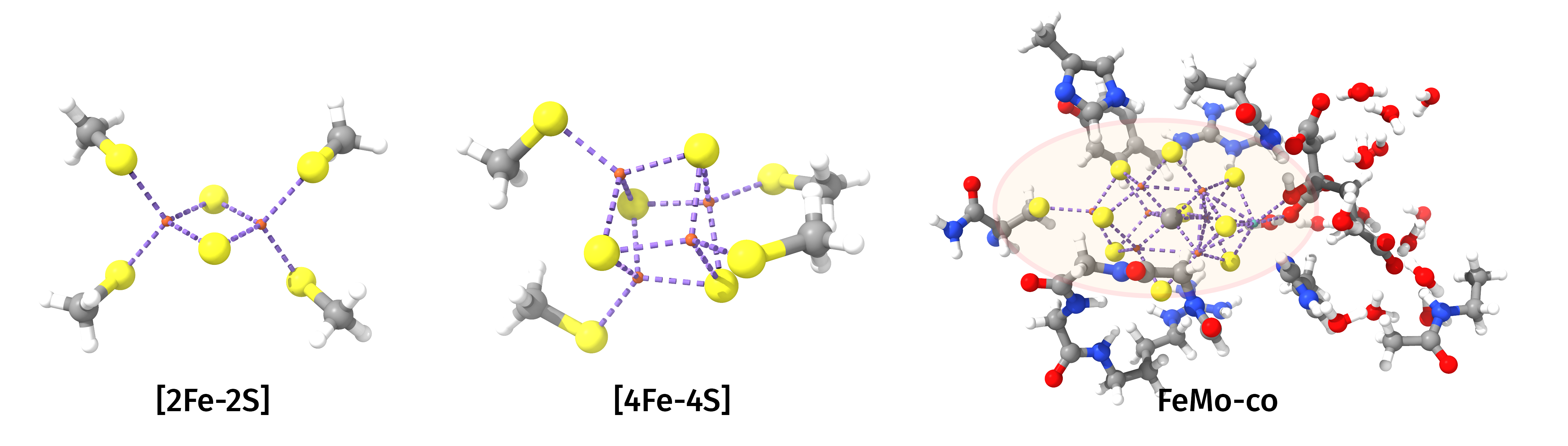}
    \caption{Fe-S clusters studied in this work: [2Fe-2S], [4Fe-4S], and the FeMo cofactor. For each cluster, we consider the ground
    state of active space Hamiltonians (see text) whose electronic structure qualitatively captures the features of the low-lying spectrum of the full \emph{ab initio} description.}
    \label{fig:fe-s}
\end{figure}

Motivated by these developments, this work has two aims. The first is to assess the performance of phaseless AFQMC on a set of Fe-S active space models\cite{li2017spin,li2019electronic} of increasing complexity ([2Fe-2S], [4Fe-4S], and the FeMo cofactor; see Figure \ref{fig:fe-s}). The second is to investigate the components that determine the method's accuracy for these systems, particularly the role of the trial wave function and the resulting walker wave function.

In phaseless AFQMC, an ensemble of non-orthogonal Slater determinants (walkers) is propagated stochastically in imaginary time, subject to a constraint, the phaseless approximation, that depends on a guess for the ground-state wave function called the trial state. The trial enters the calculation in two distinct ways: (i) it biases the walker dynamics through the phaseless constraint, and (ii) it defines the bra state in mixed estimators used to evaluate observables. Consequently, improving the trial wave function typically improves the accuracy of
the estimated ground state energies.\cite{motta2018ab, lee2022twenty} As we will demonstrate, the Fe-S clusters studied here do \emph{not} follow this typical pattern. Using our recent implementations of trial states based on the CC hierarchy, up to and including quadruple excitations,\cite{mahajan2025beyond,kjonstad2025} as well as using multi-Slater trials based on an accurate DMRG wavefunction, we show that systematically improving the trial does not systematically improve the energy. This motivates a detailed investigation of the fidelities of both trial states and walker wave functions, and of the role of different energy estimators, which taken together document some of the underlying mechanisms that limit the accuracy of phaseless AFQMC in these systems.

This paper is structured as follows. In Section \ref{sec:theory}, we review theory: phaseless AFQMC, the use of trial states based on CC theory, energy estimators, and formulae for the estimation of fidelities for both trial states and walker wave functions. This is followed in Section \ref{sec:results} by our results and a discussion for each of the Fe-S clusters, presented in order of increasing complexity ([2Fe-2S], [4Fe-4S], and the FeMo cofactor). In Section \ref{sec:conclusions} we summarize and provide some concluding remarks.

\section{Theory} \label{sec:theory}
\subsection{Phaseless approximation}

In phaseless AFQMC, an ensemble of Slater determinants or walkers is stochastically propagated in imaginary time $\tau$ by repeatedly sampling the short-time propagator to time-evolve the walkers. Such propagation provides a stochastic representation of the ground state $\vert \psi_0 \rangle$ because imaginary-time propagation filters out higher-energy states:
\begin{align}
    \ket{\psi_0} \propto \lim_{\tau \rightarrow \infty} U(\tau) \ket{\phi} = \lim_{\tau \rightarrow \infty} e^{-\tau H} \ket{\phi}, \;\; \langle \phi \vert \psi_0 \rangle \neq 0.
\end{align}
For the electronic structure of a molecular system, the system can be described by the electronic Hamiltonian
\begin{align}
    H = \sum_{pq\sigma} h_{pq} \cre{p\sigma} \ani{q\sigma} + \frac{1}{2} \sum_{pqrs\sigma\tau} g_{prqs} \cre{p\sigma} \cre{q\tau} \ani{s\tau} \ani{r\sigma},
\end{align}
where $p,q,r,s$ denote molecular orbitals and $\sigma, \tau$ denote spin coordinates.
Considering the short-time propagator, we can split the propagation of the one- and two-body terms ($H_1$ and $H_2$) via Trotterization,
\begin{align}
    U(\Delta \tau) = e^{-\frac{\Delta\tau}{2} H_1} e^{-\Delta \tau H_2} e^{-\frac{\Delta\tau}{2} H_1} + O(\Delta \tau^2).
\end{align}
We can then apply a Hubbard-Stratonovich transformation to the two-body terms, which introduces a set of auxiliary fields $\{ x_\gamma \}$,\cite{stratonovich1957,hubbard1959}
\begin{align}
\begin{split}
    U(\Delta \tau)
    &= e^{-\frac{\Delta \tau}{2} H_1} \prod_\gamma \int \frac{\dd x_\gamma}{\sqrt{2 \pi}}e^{-\frac{x_\gamma^2}{2}} e^{\sqrt{\Delta \tau} x_\gamma v_\gamma} e^{-\frac{\Delta \tau}{2} H_1} + O(\Delta \tau^2),
\end{split}
\end{align}
where
\begin{align}
    v_\gamma &= i \sum_{pq\sigma} L_{pq}^\gamma \cre{p\sigma} \ani{q\sigma}, \\
    g_{prqs} &= \sum_{\gamma} L_{pr}^\gamma L_{qs}^\gamma.
\end{align}
    Since the propagator reads
\begin{align}
    U(\Delta \tau) = \int \mathrm{d} \boldsymbol{x} \, P(\boldsymbol{x}) B(\boldsymbol{x}) + O(\Delta \tau^2),
\end{align}
    one can sample a field configuration $\boldsymbol{x}'$ from the probability distribution $P(\boldsymbol{x})$ and propagate a Slater determinant (i.e., a walker) by acting on it with the complex orbital rotation operator $B(\boldsymbol{x}')$. By repeated sampling and propagation, we obtain an ensemble of random walks that represents $\vert \psi_0 \rangle$ stochastically.

In practice, importance sampling of $P(\boldsymbol{x})$ is performed with respect to an approximation of the ground state $\vert \psi_T \rangle$ called a trial state. In this importance-sampled representation, we can express the walker wave function as
\begin{align}
    \vert \psi \rangle = \sum_n \sum_k w_k^{n} \frac{\vert \phi_k^{n} \rangle}{\langle \psi_T \vert \phi_k^{n} \rangle},
    \label{eq:walker-wf}
\end{align}
where $w_k^n$ is the weight of the $k$th walker in the $n$th timestep and $\vert \phi_k^n \rangle$ is the corresponding walker. Note that this introduction of importance sampling can already introduce a bias if the nodes of the trial state do not coincide with those of the true state.\cite{li2025fokkerplanckequationgoverningdistribution}

In each timestep, the weights are updated by projecting the importance function onto the real positive axis, i.e., we impose the phaseless approximation,\cite{zhang2003quantum}
\begin{align}
    w_k^{n+1} = \vert I_k^{n} \vert \max\bigl(0, \cos \Delta \theta_k^{n} \bigr) w_k^{n},
\end{align}
where the importance function $I_k^{n}$ is defined as
\begin{align}
    I_k^{n} = \frac{\langle \psi_T \vert B(\boldsymbol{x}-\bar{\boldsymbol{x}}) \vert \phi_k^{n} \rangle}{\langle \psi_T \vert \phi_k^n \rangle} e^{\boldsymbol{x} \cdot \bar{\boldsymbol{x}} - \frac{1}{2} \bar{\boldsymbol{x}}\cdot\bar{\boldsymbol{x}}},
\end{align}
and $\Delta \theta_k^{n}$ is the complex phase of the ratio of overlaps,
\begin{align}
    \Delta \theta_k^{n} = \arg \Bigl( \frac{\langle \psi_T \vert B(\boldsymbol{x}-\bar{\boldsymbol{x}}) \vert \phi_k^{n} \rangle}{\langle \psi_T \vert \phi_k^n \rangle} \Bigr).
\end{align}
Note that we have introduced a force-bias shift in the auxiliary fields $\bar{\boldsymbol{x}}$ that stabilizes the propagation by eliminating fluctuations to first order in $\sqrt{\Delta \tau}$ (where $\Delta \tau$ is the time step) in the importance function,\cite{motta2018ab}
\begin{align}
    \bar{x}_\gamma = - \sqrt{\Delta \tau} \frac{\langle \psi_T \vert v_\gamma \vert \phi_k^n \rangle}{\langle \psi_T \vert \phi_k^n \rangle}.
\end{align}

The combination of a force bias shift and real-positive weight-projection overcomes the fermion sign problem and its associated exponential cost, provided that $\langle \psi_T \vert \phi \rangle$ can be evaluated in polynomial time for the given $\vert \psi_T \rangle$. However, it simultaneously introduces a further bias in the walkers whose magnitude depends on $\vert \psi_T \rangle$.\cite{zhang2003quantum,motta2018ab} In general, this means the walker wave function in Eq.~\eqref{eq:walker-wf} no longer converges to the true ground state $\vert \psi_0 \rangle$.

\subsection{Ground state energy estimation} \label{sec:energy-estimation}
Given a trial $\langle \psi_T \vert$ and walker wave function $\vert \psi \rangle$,
the ground state energy is typically estimated using the mixed-energy estimator
\begin{align}
    E_0 = \frac{\langle \psi_T \vert H \vert \psi \rangle}{\langle \psi_T \vert \psi \rangle} = \frac{\sum_I w_I E_L(\phi_I)}{\sum_I w_I}, \label{eq:afqmc-energy}
\end{align}
where $I = (k,n)$ and where we have defined the \emph{local} energy
\begin{align}
    E_L(\phi_I) = \frac{\langle \psi_T \vert H \vert \phi_I \rangle}{\langle \psi_T \vert \phi_I \rangle}.
\end{align}
When evaluating the energy according to Eq.~\eqref{eq:afqmc-energy}, one typically projects the local energy onto the real axis.

Although not standard, one can also estimate the energy with a different trial than the one used for the importance sampling and phaseless constraint. In this context, we will distinguish the \emph{measurement} trial $\langle \psi_T \vert$ (used for energy evaluation) from the \emph{guiding} trial $\langle \psi_G \vert$ (used for importance sampling and projection of weights). Then, the energy estimator reads
\begin{align}
    E_0 = \frac{\langle \psi_T \vert H \vert \psi \rangle}{\langle \psi_T \vert \psi \rangle} = \frac{\sum_I w_I F(\phi_I) E_L(\phi_I)}{\sum_I w_I F(\phi_I)}, \label{eq:guide-trial-energy}
\end{align}
where we have introduced the \emph{reweighting factors}
\begin{align}
    F(\phi_I) = \frac{\langle \psi_T \vert \phi_I \rangle}{\langle \psi_G \vert \phi_I \rangle}. \label{eq:reweighting}
\end{align}
As for Eq.~\eqref{eq:afqmc-energy}, we project onto the real axis when evaluating local quantities.

The estimators in Eqs.~\eqref{eq:afqmc-energy} and \eqref{eq:guide-trial-energy} have some properties worth pointing out. For both estimators, the exact ground state energy is reproduced if the measurement trial becomes exact, $\vert \psi_T \rangle = \vert \psi_0 \rangle$, and this is true independently of the walker wave function. This is the well-known result that the phaseless AFQMC method predicts the exact ground state energy when the (measurement) trial becomes exact.\cite{motta2018ab}
Note that this is separate from the question of whether the walker wave function becomes exact in this limit.\cite{li2025fokkerplanckequationgoverningdistribution}
Moreover, both estimators are non-variational, implying that improvements in the quality of $\vert \psi_T \rangle$ or $\vert \psi \rangle$, or even both, are not guaranteed to reduce errors.

The estimators also have important differences. The estimator in Eq.~\eqref{eq:guide-trial-energy} can in general become more noisy than the one in Eq.~\eqref{eq:afqmc-energy}, as the effective weights $w_I F(\phi_I)$ vary more strongly than $w_I$.
However, for the systems considered in this work, we find for both estimators that the stochastic errors (for a given number of samples) mainly depend on the choice of measurement trial.

An interesting difference between the estimators occurs in the limit where the guiding trial is more accurate than the measurement trial. At first glance, it would seem to be counter-productive to choose not to make use of a more accurate trial for measuring the energy. However, the situation is more nuanced since the measurement trial can serve the purpose of effectively filtering out certain excitations in the walker wave function $\vert \psi \rangle$. To see this, consider the example where the measurement wave function $\vert \psi_T \rangle$ is the HF state. Then, from rank-considerations, the energy estimate becomes exact provided the walker wave function $\vert \psi \rangle$ is exact in the singles and doubles subspace. More generally, to obtain the exact ground state energy, $\vert \psi \rangle$ needs to be exact only in the $r_T + 2$ excitation subspace, where $r_T$ is the rank of the highest non-zero excitation in the measurement trial $\vert \psi_T \rangle$. Interestingly, this effectively places less strict requirements on the accuracy of higher-order terms in the wave function $\vert \psi \rangle$ when estimating the energy. As discussed further below, this suggests that using a \emph{less} accurate measurement trial may in some cases lead to a more accurate estimate of the ground state energy.
Note that this is not restricted to phaseless AFQMC.
A similar behavior occurs in CC theory, where the bra state in the energy expression is the reference state and, thus, only the reference, singles and doubles components of the CC wave function contribute to the energy.

\subsection{Trial states based on coupled cluster theory}
Coupled cluster theory is considered one of the most accurate approaches for capturing dynamical correlation in the ground state,\cite{bartlett2007coupled} making its use as a trial state of particular interest. In this method, the  ground state has an exponential form,\cite{bartlett2007coupled}
\begin{align}
    \vert \mathrm{CC} \rangle = e^T \vert \phi_0 \rangle, \quad T = \sum_{i=1}^n T_i,
\end{align}
where the cluster operator $T$ generates excitations out of the reference $\vert \phi_0 \rangle$ (typically a HF state) and is truncated at an excitation order $n$ (e.g., $n=2$ corresponds to CCSD\cite{purvis1982full}). The parameters in $T$, the individual amplitudes that weight the different excitations, are determined from a set of equations
obtained by projecting the Schrödinger equation onto this same excitation subspace.\cite{helgaker2013molecular}

This exponential form of $\vert \mathrm{CC} \rangle$ poses a challenge since $\langle \mathrm{CC} \vert \phi \rangle$ has a cost that scales exponentially with the size of the system. Additional approximations are therefore usually made when applying $\vert \mathrm{CC} \rangle$ as a trial state.
One simple approach, which we adopt in this work, is to project $\vert \mathrm{CC} \rangle$ into a configuration interaction (CI) subspace, instead defining the trial state as\cite{mahajan2025beyond,kjonstad2025}
\begin{align}
    \vert \psi_T \rangle = \mathcal{P} \vert \mathrm{CC} \rangle, \quad \mathcal{P} = \vert \phi_0 \rangle\langle \phi_0 \vert + \sum_{i = 1}^n \mathcal{P}_i,
\end{align}
where $\mathcal{P}_i$ is a projection operator onto the subspace of $i$-fold excitations relative to the reference state $\vert \phi_0 \rangle$. This projection ensures polynomial scaling, providing a hierarchy of methods with increased cost and (in typical cases) accuracy, but it also has well-known limitations for large systems (loss of size-extensivity). The overall computational scaling with system size $N$ using these trial states mirrors the scaling to determine the trial states themselves: $O(N^6)$, $O(N^8)$, and $O(N^{10})$ for trial states that include up to double (CCSD), triple (CCSDT\cite{noga1987full}), and quadruple (CCSDTQ\cite{kucharski1992coupled}) excitations, respectively, with the sampling part of the algorithm scaling more favorably for triple and quadruple excitations, $O(N^7)$, and $O(N^{9})$. We refer to the original works on these trial states for more details regarding the algorithm and implementation, along with benchmark results on molecular systems.\cite{mahajan2025beyond,kjonstad2025}

Note that any type of trial state can in principle be applied in the  phaseless AFQMC method.
In this work, we will use both HF and multi-Slater determinant (MSD) trials, in addition to the
CC trials described above.
The reader is referred to the literature for more details
regarding HF and MSD trial states.\cite{zhang2003quantum,landinez2019non,mahajan2022selected}

\subsection{Fidelity estimation for trial states and walker wave functions}
Understanding the accuracy of phaseless AFQMC is far from straight-forward,
given the complex relationship between the trial and walker wave functions, and
their influence on the accuracy of properties
like the ground state energy $E_0$
(see Section \ref{sec:energy-estimation}).
In this work, we use a combination of direct measures and indirect measures to
better understand the performance of the method.

The most direct measure of quality is given by estimating the \emph{fidelity}
of the relevant approximate wave functions with
respect to some exact or near-exact reference for the ground state $\vert \psi_0 \rangle$.
We define the fidelity of some state $\vert \chi \rangle$ relative to $\vert \psi_0 \rangle$ as
\begin{align}
    f(\chi) = \frac{\vert \langle \chi \vert \psi_0 \rangle \vert}{\langle \chi \vert \chi \rangle^{1/2} \langle \psi_0 \vert \psi_0 \rangle^{1/2}}.
\end{align}
In the
context of phaseless AFQMC, we are concerned both with the fidelity of the trial $\vert \psi_T \rangle$
and the walker wave function $\vert \psi \rangle$. Taken together, these states determine
the accuracy of observables like the energy $E_0$ and their fidelities may serve to explain
observed errors. However, one should note that improvements
in $\vert \psi_T \rangle$ and $\vert \psi \rangle$ fidelity may not necessarily translate into improvements in the prediction of $E_0$.
This stems from the non-variational form of the energy estimator, see Eq.~\eqref{eq:afqmc-energy}.

For the systems studied in this work, we determine the reference state $\vert \psi_0 \rangle$
either by full configuration interaction (FCI) or via the
density-matrix renormalization group (DMRG).\cite{schollwock2005density} In the case of DMRG,
fidelities are evaluated by converting the matrix product state (MPS) to CI form i.e.,
\begin{align}
\begin{split}
    \vert \psi_0 \rangle &= \sum_{n_1 n_2 \ldots n_K} \sum_{\{ \alpha_k \}} A_{\alpha_1}^{n_1}[1] A_{\alpha_1\alpha_2}^{n_2}[2]\cdots A_{\alpha_{K-1}}^{n_K}[K] \vert n_1 n_2 \cdots n_K \rangle \\
    &= \sum_I c_I \vert I \rangle,
\end{split}
\end{align}
either with some
coefficient cutoff $\tau$,~\cite{lee2021externally}
or by sampling CI configurations $\vert I \rangle$ from the MPS $\vert \psi_0 \rangle$ with
probability given by $p_I \sim |c_I|^2$.\cite{guo2018communication}

Once the reference $\vert \psi_0 \rangle$ is expressed in CI form,
we can readily evaluate the fidelity $f(\psi_T)$ of the (CI) trial state $\vert \psi_T \rangle$.
For the special case where $\vert \psi_0 \rangle$ and $\vert \psi_T \rangle$ are
 not expressed in the same orbital basis,
we evaluate $\langle \psi_T \vert \psi_0 \rangle$ by making use of the overlap implementation for walkers; i.e.,
we expand the overlap as
$\sum_I c_I \langle \psi_T \vert I \rangle$
and make use of the $\langle \psi_T \vert \phi \rangle$ implementation with
the ``walker'' set to
$\vert \phi \rangle = \vert I \rangle$. Finally, in the case where
configurations are sampled from $\vert \psi_0 \rangle$ with probability $p_I \sim c_I^2$,
we use the relation
\begin{align}
    \langle \psi_T \vert \psi_0 \rangle = \sum_I c_I^2 \Bigl( \frac{\langle \psi_T \vert I \rangle}{c_I} \Bigr) = \sum_I p_I O_I
\end{align}
and, hence, the overlap can be estimated as
\begin{align}
    \langle \psi_T \vert \psi_0 \rangle \approx \frac{1}{M} \sum_{I_k \sim p_I} O_{I_k},
\end{align}
where we have drawn $M$ samples $\{ I_k \}$ from $p_I$. The variance is similarly estimated by the
corresponding unbiased variance estimator.

Evaluating the fidelity $f(\psi)$ of the walker wave function $\vert \psi \rangle$ is
more involved. Using Eq.~\eqref{eq:walker-wf},
we see that, for a given number of walker samples $\{ I\}$, the fidelity of the walker wave function can be expressed as
\begin{align}
    f(\psi) = \frac{\vert\langle \psi_0 \vert \psi \rangle\vert}{\langle \psi_0 \vert \psi_0 \rangle^{1/2}\langle \psi \vert \psi \rangle^{1/2}},
\end{align}
where
\begin{align}
    \langle \psi_0 \vert \psi \rangle = \sum_I w_I \frac{\langle \psi_0 \vert \phi_I \rangle}{\langle \psi_T \vert \phi_I \rangle} \label{eq:numerator}
\end{align}
and
\begin{align}
     \langle \psi \vert \psi \rangle = \sum_{IJ} \frac{w_I \langle \phi_I \vert \phi_J \rangle w_J}{\langle \psi_T \vert \phi_I \rangle^\ast \langle \psi_T \vert \phi_J \rangle}. \label{eq:denominator}
\end{align}

Note that $f(\psi)$ is an approximation of the fidelity of the walker distribution whose accuracy
depends on the number of samples drawn $\{ I \}$.
In our case, the number of $\{ I \}$ is equal to the number of blocks $\times$ number of walkers,
as samples are stored at the end of each block (composed of a set of timesteps)
 and not in every individual timestep.

 In practice, determining a reliable estimate of $f(\psi)$ requires a large
 number of walker samples.
 For reduced models of the [2Fe-2S] system, we find that we require on the
 order of $5\,000 \times 200 = 10^6$ samples to converge the fidelity,
 which translates to $10^{12}$ number of $\langle \phi_I \vert \phi_J \rangle$ evaluations
 to obtain $\langle \psi \vert \psi \rangle$, each of which scales as $O(N^3)$,
 where $N$ is the system size.
 Additional approximations are thus required to evaluate
 Eqs.~\eqref{eq:numerator} and \eqref{eq:denominator}.
 To evaluate these terms approximately, we estimate each quantity via importance sampling, where
 for Eq.~\eqref{eq:numerator} we draw a set of samples $\{ I_k \}$ from
 \begin{align}
    p_I = \frac{w_I}{\vert \langle \psi_T \vert \phi_I \rangle \vert} \Big/ \sum_J \frac{w_J}{\vert \langle \psi_T \vert \phi_J \rangle \vert}.
 \end{align}
 For Eq.~\eqref{eq:denominator}, we draw samples $\{ I_k, J_k \}$ by drawing $I$ and $J$
 independently from the same distribution, effectively sampling from the product distribution $p_{IJ} = p_I p_J$.
Additionally, when evaluating Eq.~\eqref{eq:numerator}, we perform local real-axis projection of individual
terms. As for the energy estimators, we find in test cases that the overall complex phase is zero to within
stochastic error and does not significantly
affect the resulting fidelity estimate.

Note that with this importance sampling procedure,
there are two sources of error in $f(\psi)$: (a) the finite number of
walker samples extracted from the imaginary-time propagation; and (b) the finite number of sampled
indices $\{I_k\}$ and $\{I_k, J_k\}$ used to evaluate Eqs.~\eqref{eq:numerator} and \eqref{eq:denominator}
from a given sample of walkers $\{ I \}$.
For (b), we estimate the uncertainty via standard variance estimators
for Eqs.~\eqref{eq:numerator} and \eqref{eq:denominator}
and apply the delta method to estimate the resulting error in $f(\psi)$.
To estimate errors due to (a), we partition the walker samples $\{ I \}$ into $K$
non-overlapping windows and estimate the variance of fidelity estimates across the windows,
subtracting the average variance due to importance sampling of Eqs.~\eqref{eq:numerator}
and \eqref{eq:denominator}.
This procedure requires that each window contains enough blocks for the
within-window fidelity estimate to be approximately converged, such that
finite-sample bias does not vary across windows and artificially inflate the inter-window
variance.
Based on cumulative convergence plots (see Supporting Information S1), we find that the fidelity
estimate converges on a scale of
about $5\,000$ blocks for the systems studied in this work; we therefore use $K = 10$ windows
to estimate variance due to (a).
The overall uncertainty in $f(\psi)$ is then obtained
by adding the variances from (a) and (b).
See Supporting Information S1 for more details on the estimation procedure for $f(\psi)$.

\section{Results and discussion} \label{sec:results}
We consider active space models for three Fe-S clusters (Figure \ref{fig:fe-s}):
the [2Fe-2S] and [4Fe-4S] models reported by Li and Chan\cite{li2017spin} and
the FeMo cofactor model reported by Li, Li, Dattani, Umrigar, and Chan.\cite{li2019electronic} These
active spaces are constructed from the Fe and Mo $3d$/$4d$ orbitals and ligand
$2s$/$2p$ and $3p$
orbitals on C and S, and are designed to capture the main spin
and charge physics of the low-lying spectrum.\cite{li2017spin,li2019electronic}
All three clusters exhibit a dense manifold of low-lying states. This manifold arises from
competing spin and charge configurations on the Fe centers,
which formally adopt oxidation states (II) and (III)
(corresponding to $d^6$ and $d^5$ and $S=2$
and $S=5/2$, respectively), and, in the case of the FeMo cofactor,
a formal (III) oxidation state on Mo ($d^3$ and $S=1/2$).\cite{kowalska2019x,bjornsson2014identification}
The local spin moments of the Fe centers reflect Hund's rule applied to the open-shell $d$ electrons,
but the low-energy spectrum is further shaped by competing double-exchange and
superexchange interactions between the centers.

Such spin and charge polarization can be qualitatively captured by broken-symmetry (BS)
states at the mean-field level, e.g.~using unrestricted Hartree-Fock (UHF) or
unrestricted density functional theory,\cite{lovell2001femo} and these references
serve as effective starting points for correlated treatments. We therefore
enumerate BS-UHF states by the distribution of $\alpha$- and $\beta$-spin density across
the Fe centers, yielding distinct spin and charge isomers, and group these into
families by spatial spin arrangement (denoted BS$n$ with $n = 0, 1, \ldots$).
This procedure yields, in general, a number of UHF references, to each of
which we then apply correlated methods (CC, AFQMC, and DMRG) to recover
dynamic and static correlation. An important caveat is that, because a given
BS family may have negligible overlap with the true ground state, the corresponding
correlated solutions may not approximate it; they may instead track an excited state.
The full set of BS references must therefore be considered, with the ground
state being identified \emph{a posteriori}.

\subsection{Computational details}
The UHF references are converged using \textsc{PySCF}.\cite{sun2020recent,sun2026python}
All phaseless AFQMC calculations are performed with AD-AFQMC,\cite{ad_afqmc_software} where we use recent implementations of CCSD, CCSDT, and CCSDTQ trial states.\cite{mahajan2025beyond,kjonstad2025} Unrestricted CC trial states are converged either with \textsc{PySCF} (CCSD) or CCPY (CCSD, CCSDT, CCSDTQ).\cite{gururangan2024ccpy}
Throughout, we use unrestricted CC trial states, and when we write, e.g., ``CCSD,'' we mean unrestricted CCSD (and similarly for higher-order CC trial states).
We use spin-unrestricted walkers (200 for [2Fe-2S] and [4Fe-4S]; 100 for the FeMo cofactor),
a timestep of $0.005$ a.u., and Cholesky thresholds $\leq 10^{-6}$, and we propagate for a variable number of blocks either to reach a desired stochastic error in the energy or to sufficiently sample the walkers to reliably estimate fidelities. All DMRG calculations were performed using \textsc{block2}.\cite{zhai2023block2} In these calculations, we either perform spin-adapted calculations (SA-DMRG) or spin-unrestricted calculations using broken-symmetry orbitals (U-DMRG). For AFQMC calculations based on DMRG and FCI references, we use the existing MSD implementation in AD-AFQMC. Trial state and walker wave function fidelities are estimated by making use of the existing implementations of overlaps between walkers and CC and MSD trial states in the AD-AFQMC package.

\subsection{The [2Fe-2S] dimer}
We restrict ourselves to the oxidized dimer, \ce{[Fe2S2(SCH3)4]}$^{2-}$,
where both Fe atoms are in the formal oxidation state (III).
The active space model by Li and Chan\cite{li2017spin} consists of 30 electrons in 20 orbitals,
and includes $3d$ Fe orbitals and $3p$ S orbitals.
The ground state of this dimer is known to be an overall singlet ($S=0$) arising
from anti-ferromagnetically coupled Fe(III) centers ($d^5$)
stabilized by the bridging S orbitals.\cite{li2017spin,sharma2014low}
As a result, there is a single spin-isomer corresponding to five $\alpha$
electrons on one Fe atom and five $\beta$ electrons on the other Fe,
leading to a single BS-UHF solution that serves
as the starting point for correlated treatments.

\begin{figure}[!htbp]
    \centering
    \includegraphics[width=.9\linewidth]{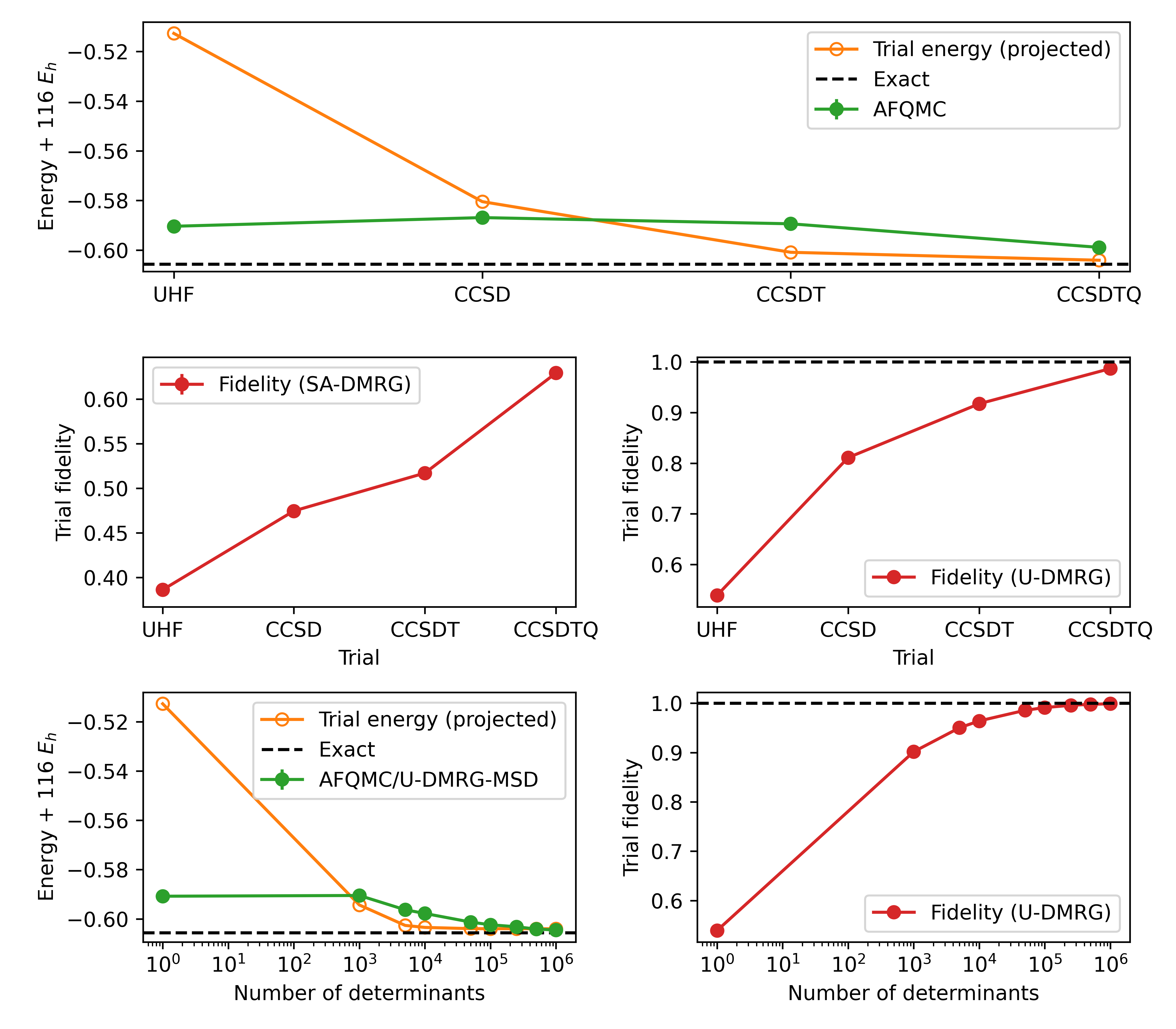}
    \caption{Energies and trial fidelities for [2Fe-2S] (30e, 20o). Upper: Energies of CC trial states and associated AFQMC energies. Middle: Fidelity of trial states relative to (left) SA-DMRG ($D=1000$, $E_0 = -116.60560$ $E_h$) and (right) U-DMRG ($D=5000$, $E_0 = -116.60406$ $E_h$) references. Lower left: Energies of MSD trial based on U-DMRG determinants and associated AFQMC energies. Lower right: Fidelity of the U-DMRG MSD trial state relative to U-DMRG reference. Fidelities relative to U-DMRG are evaluated with a truncated expansion with total weight $\sum_I c_I^2 = 0.994$.}

    \label{fig:fe2-30e-20o-afqmc-std}
\end{figure}

Our findings for the [2Fe-2S] (30e, 20o) model are given in Figure \ref{fig:fe2-30e-20o-afqmc-std}. Considering the AFQMC energies using CC trials (upper panel), we see that the energies do not improve when going from a UHF trial to a CCSD or a CCSDT trial, with predicted energies even becoming slightly less accurate. More surprisingly, we see an ``inverted'' region (for CCSDT and CCSDTQ trial states) where the AFQMC energies are \emph{less} accurate than the projected energies of the trial on which they are based. This stands in contrast to the typical behavior of phaseless AFQMC, which is usually found to significantly improve upon the energy of the underlying trial state. This is also not a matter of the type of trial state, as our recent benchmarks on simpler molecular systems show that CC trial states typically lead to more accurate phaseless AFQMC energies.\cite{mahajan2025beyond,kjonstad2025}

Given these results, one might question whether the improvement in CC energies, as one climbs the hierarchy, coincides with a real improvement in the trial state fidelity.
Indeed, low-fidelity trial states at the CCSD and CCSDT levels would offer a straight-forward explanation of the observed inverted region.
However, the opposite appears to be the case: upon evaluating the fidelity of the CC trial states against both spin-adapted and unrestricted DMRG references, we find that the fidelities of the trial states are in fact \emph{increasing} as we ascend the CC hierarchy (Figure \ref{fig:fe2-30e-20o-afqmc-std}, center). This points to a non-trivial relationship between the trial state fidelity and the errors of AFQMC for this system. Examples where higher-fidelity trials do not provide more accurate energies have also been observed in some previous works.\cite{kjonstad2025, amsler2023classical}

A related question is whether the wrong components are added to the trial state as one ascends the CC hierarchy, in which case the overall fidelity would increase while important components of the ground state were nevertheless neglected. However, we reproduce the same inverted region when we incorporate the most important configurations extracted from a DMRG reference state (unrestricted DMRG, $D=5000$) and use an MSD trial state (see Figure \ref{fig:fe2-30e-20o-afqmc-std}, lower panel). To ensure a proper comparison with the CC trial state case, we compare the phaseless AFQMC/MSD energies with the \emph{projected} MSD trial energy
\begin{align}
    E_0 = \frac{\langle \phi_0 \vert H \vert \psi_T \rangle}{\langle \phi_0 \vert \psi_T \rangle}.
\end{align}
Note that this projected energy may well be more accurate than the variational energy of the MSD trial, in the same way that the CC energy may well be more accurate than the variational energy of the CC state (and the CI-projected CC trial state). In particular, for a sufficiently accurate DMRG reference, the projected energy becomes exact at the point when all configurations in the CI singles and doubles (CISD) subspace are included in the MSD trial. Similarly, the CC energy becomes exact when its projection into the CISD subspace becomes exact.

Another possibility is that the walker wave function fidelities may be decreasing even as the trial state fidelities are increasing; there is no guarantee that a more accurate trial will lower the bias of the walker ensemble due to the phaseless projection. To investigate this possibility, we have compressed the (30e, 20o) active space model by Li and Chan\cite{li2017spin} into a smaller (18e, 14o) active space model that preserves the 10 Fe $d$ orbitals along with $4$ bridging orbitals associated with S. To construct this active space, we use the natural orbital basis of a near-exact spin-adapted DMRG reference ($D=8000$) and freeze orbitals that are largely doubly occupied. For the double-occupancy threshold, we find that a value of $n_o = 1.97$ yields the smallest active space that reproduces our findings for the (30e, 20o) active space. For more details on the active space construction, we refer to Supporting Information S2. The integral files for the (18e, 14o) active space are also provided in a separate paper data repository.\cite{zenodo}

\begin{figure}[!htbp]
    \centering
    \includegraphics[width=0.8\linewidth]{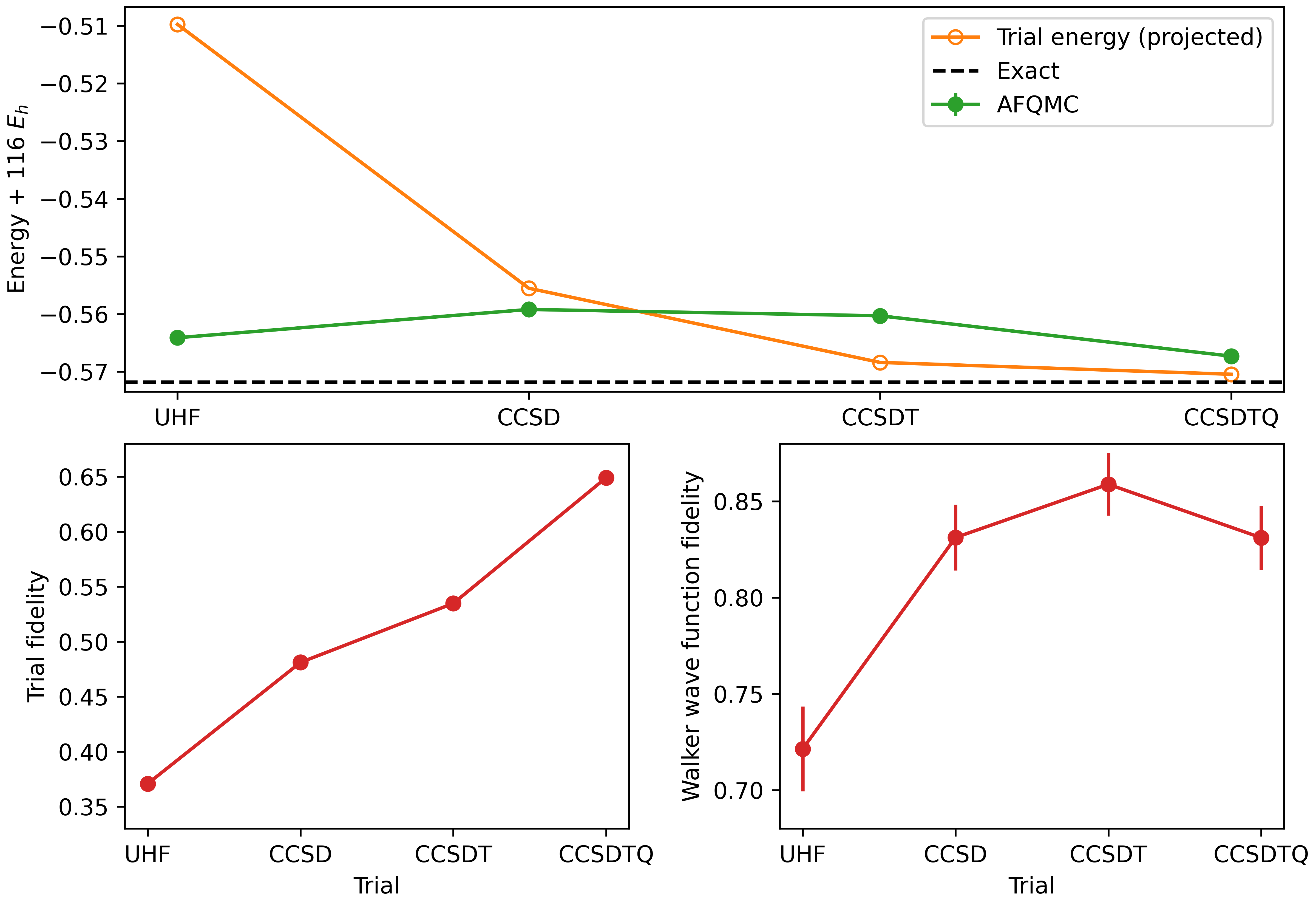}
    \caption{Energies, trial fidelities, and walker wave function fidelities for [2Fe-2S] (18e, 14o). Upper: Energies of CC trial states and associated AFQMC energies. Lower: Fidelity of trial states (left) relative to the exact FCI reference and fidelities of walker wave functions (right) relative to an FCI/MSD reference state with total weight $\sum_I c_I^2 = 0.995$.}
    \label{fig:fe2-18e-14o-afqmc-std}
\end{figure}

This smaller (18e, 14o) active space model for [2Fe-2S] makes it feasible to estimate the walker wave function fidelities via exhaustive sampling ($10^6$ samples). The results are shown in Figure \ref{fig:fe2-18e-14o-afqmc-std}.
First considering the energies (upper panel), we see by comparison to Figure \ref{fig:fe2-30e-20o-afqmc-std} that the trends in the observed errors in the CC and AFQMC energies are nearly identical, including the presence of the same inverted region where the trial state energy is more accurate than the phaseless AFQMC energy. Furthermore, the trial state fidelities (lower left) are  nearly identical across the two models (compare Figure \ref{fig:fe2-30e-20o-afqmc-std}, middle left). For the walker wave function fidelities, we find that they are generally higher than the trial state fidelities, and that higher walker fidelities are obtained with higher-fidelity trial states (Figure \ref{fig:fe2-18e-14o-afqmc-std}, lower left and right). Note that the walker fidelity is improved significantly for a CCSD trial state when compared to the UHF trial, but there is no noticeable further increase  for CCSDT and CCSDTQ trials. The comparatively higher fidelity of the walker wave functions, relative to the spin-broken trial states, likely arises because the spin-unrestricted walkers become restricted singlets in the long-time limit (see Supporting Information S1).
The perhaps more surprising finding here is that the AFQMC energies are becoming less accurate even as \emph{both} the trial state and the walker wave functions are improving significantly.
However, since the energy estimator is non-variational, one possible explanation is that UHF trial states benefit from a favorable cancellation of errors for this system.

\begin{figure}[!htbp]
    \includegraphics[width=\linewidth]{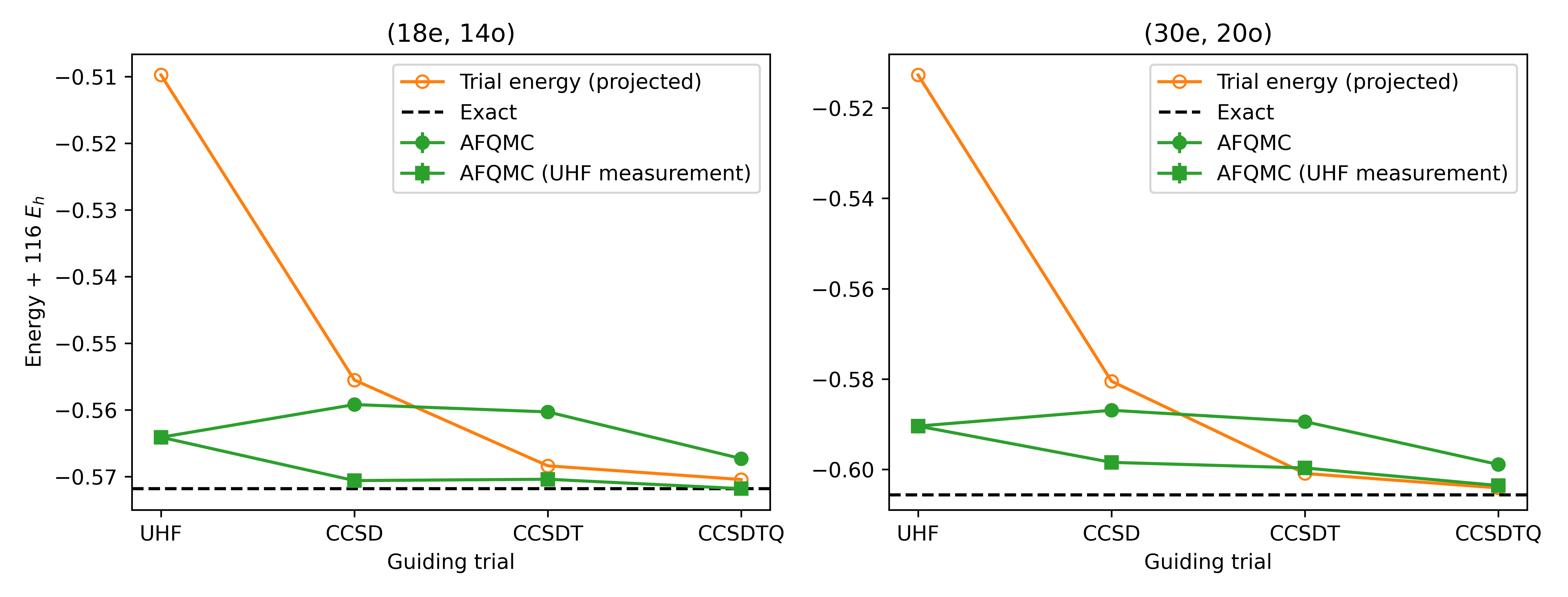}
    \caption{Energies for the [2Fe-2S] dimer including AFQMC energies with the standard estimator (filled circles) and with the measurement trial given by UHF (filled squares). Left: (18e, 14o) active space model. Right: (30e, 20o) active space model.}
    \label{fig:fe2-uhf-measurement}
\end{figure}

One source of such cancellation may be related to the particular form of the standard energy estimator, Eq.~\eqref{eq:afqmc-energy}, which exposes higher-order excitation sectors of the walker wave function when the trial is improved by adding contributions to higher-order sectors. For example, the estimator exposes up to quadruple excitations in the walkers when using a CCSD trial state (recall that this state is projected and inhabits the CISD subspace). In contrast, for a UHF trial state, only walker components in the CISD subspace are exposed and contribute to the energy. This can naturally affect the accuracy of the energies if the walker fidelities are high in low-order excitation sectors but deteriorate in quality in higher-order sectors. An indirect way to investigate this question is to separate the guide from the measurement trial and use the UHF reference $\langle \phi_0 \vert$ in the energy measurement, see Eq.~\eqref{eq:guide-trial-energy}. With this choice, new higher-order excitation sectors in the walkers are not exposed as one ascends the CC hierarchy. Energies obtained for this guide/measurement separation are shown in Figure \ref{fig:fe2-uhf-measurement} (see also Supporting Information S5 for comparisons of local energy variances). As is clear from these results, this choice improves the accuracy of the phaseless AFQMC energies significantly. However, as we will see, the findings for this choice are more mixed for the larger, more complex Fe-S clusters.

\begin{figure}[!htbp]
    \includegraphics[width=\linewidth]{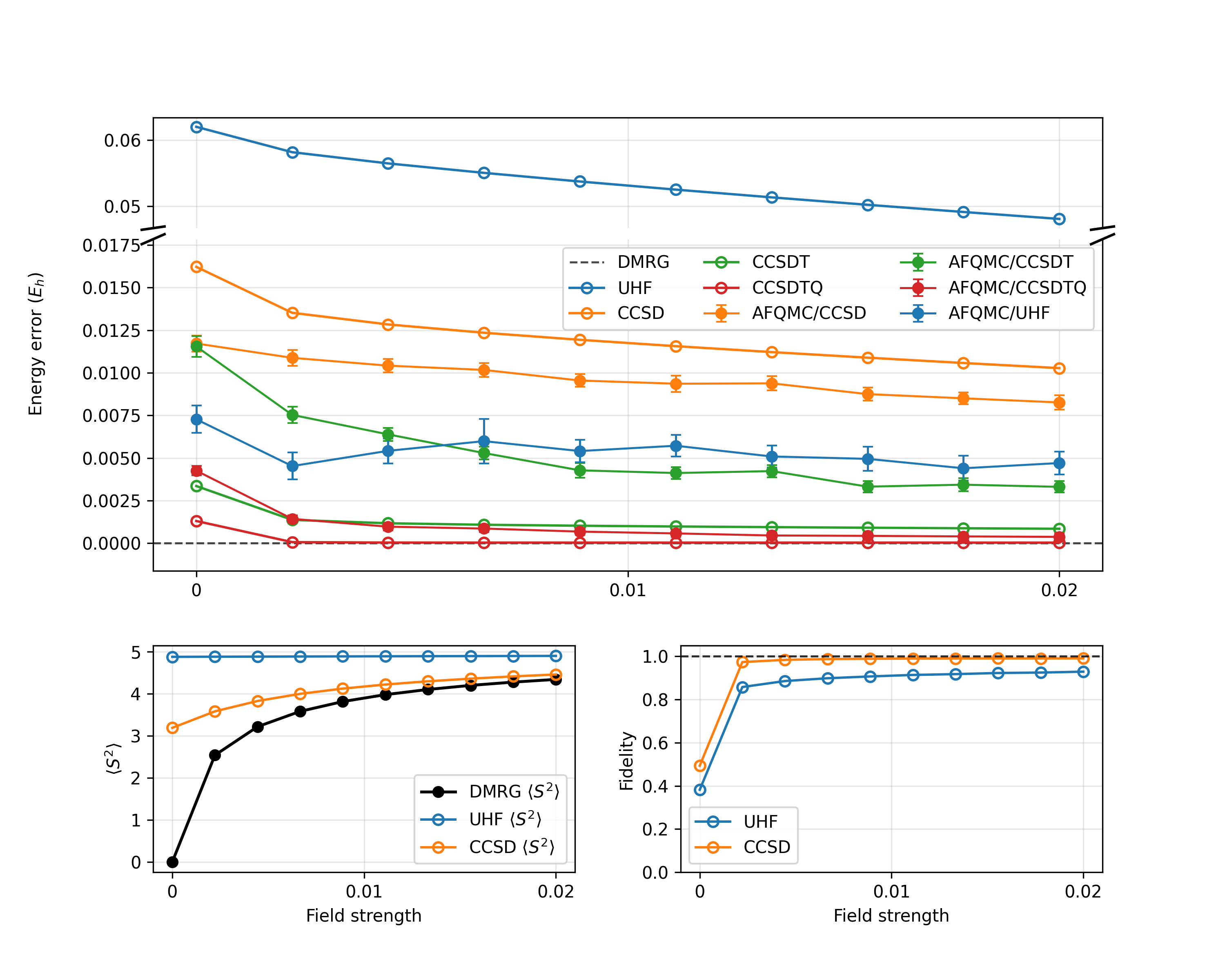}
    \caption{Energies, $\langle S^2 \rangle$, and trial state fidelities for [2Fe-2S] (18e, 14o) with a symmetry-breaking spin-Zeeman field that stabilizes the Fe$_{A}(\uparrow)$--Fe$_{B}(\downarrow)$ broken-symmetry solution (see text). Field strengths are in a.u. Upper panel: trial state and AFQMC energy errors relative to  DMRG ($D=3000$) references. Lower left: $\langle S^2 \rangle$ values computed with UHF, CCSD, and DMRG. Lower right: Trial state fidelities computed relative to the DMRG reference via sampling of configurations.}
    \label{fig:fe2-spin-breaking-field}
\end{figure}

One may also wonder whether these findings are an artifact of using spin-unrestricted trial states and walkers. The ground state is a singlet, yet the trial states are significantly spin-contaminated, and this mismatch could plausibly worsen the bias induced by the phaseless constraint. To test this, we apply a fictitious staggered local spin-Zeeman field that breaks the $S^2$ symmetry of the Hamiltonian and stabilizes the broken-symmetry solution. Specifically, we add the one-body term
\begin{align}
    H_f = - 2 \lambda (S_A^z - S_B^z), \quad S_X^z = \frac{1}{2} \sum_{p \in X} (n_p^\alpha - n_p^\beta),
\end{align}
where $\lambda$ is the field strength and $S_X^z$ is the $z$-projection of the spin  on the Fe$_X$ center. Here, $p \in X$ if $p$ is a $d$-orbital on the Fe$_X$ center. In effect, the field given by $H_f$ shifts the $\alpha$ and $\beta$ orbital energies in opposite directions on each Fe center, with the pattern reversed between Fe$_A$ and Fe$_B$, thereby stabilizing the Fe$_{A}(\uparrow)$--Fe$_{B}(\downarrow)$ broken-symmetry solution for $\lambda > 0$.

Scanning $\lambda$ over values on the order of the spin-ladder spacing, we find that the zero-field energy trends are reproduced for field strengths up to
$\lambda = 0.02$ a.u.~(Figure \ref{fig:fe2-spin-breaking-field}). In particular, the inverted energy trend for CCSDT and CCSDTQ trial states persists throughout $\lambda \in [0, 0.02]$ a.u., although it becomes less pronounced for large $\lambda$. Note that the inversion persists even as the trial state fidelities approach unity ($>0.99$) and the CCSD $\langle S^2 \rangle$ converges toward the exact value.
As in the zero-field case, we find that applying a CCSD guide and a UHF measurement trial leads to higher accuracy for $\lambda \in [0, 0.02]$ a.u.~(see Supporting Information S4).
Overall, these results strongly suggest that the non-monotonic convergence is not solely an artifact of spin-unrestricted trial states, and that the relatively low trial-state fidelities in the field-free case do not, by themselves, account for the non-monotonic behavior or the low accuracy of the energies.

\subsection{The [4Fe-4S] cubane}
The cubane cluster (\ce{[Fe4S4(SCH3)4]}$^{2-}$)  active space model comprises 54 electrons in 36 orbitals, including the $3d$ Fe orbitals and $3p$ S orbitals.\cite{li2017spin} Also for this system, the ground state is an overall singlet formed by anti-ferromagnetic coupling of the four Fe centers, two of which are formally in oxidation state (II) and two in (III). By distributing the Fe oxidation and spin configurations, i.e.~$\alpha$/$\beta$ Fe(II) and $\alpha$/$\beta$ Fe(III) in four different positions, we can construct a number of physically meaningful initial spin densities. From these, we have characterized 26 distinct BS UHF references. These references have been categorized into three BS families and are given in Supporting Information S3.

\begin{figure}[!htbp]
    \includegraphics[width=\linewidth]{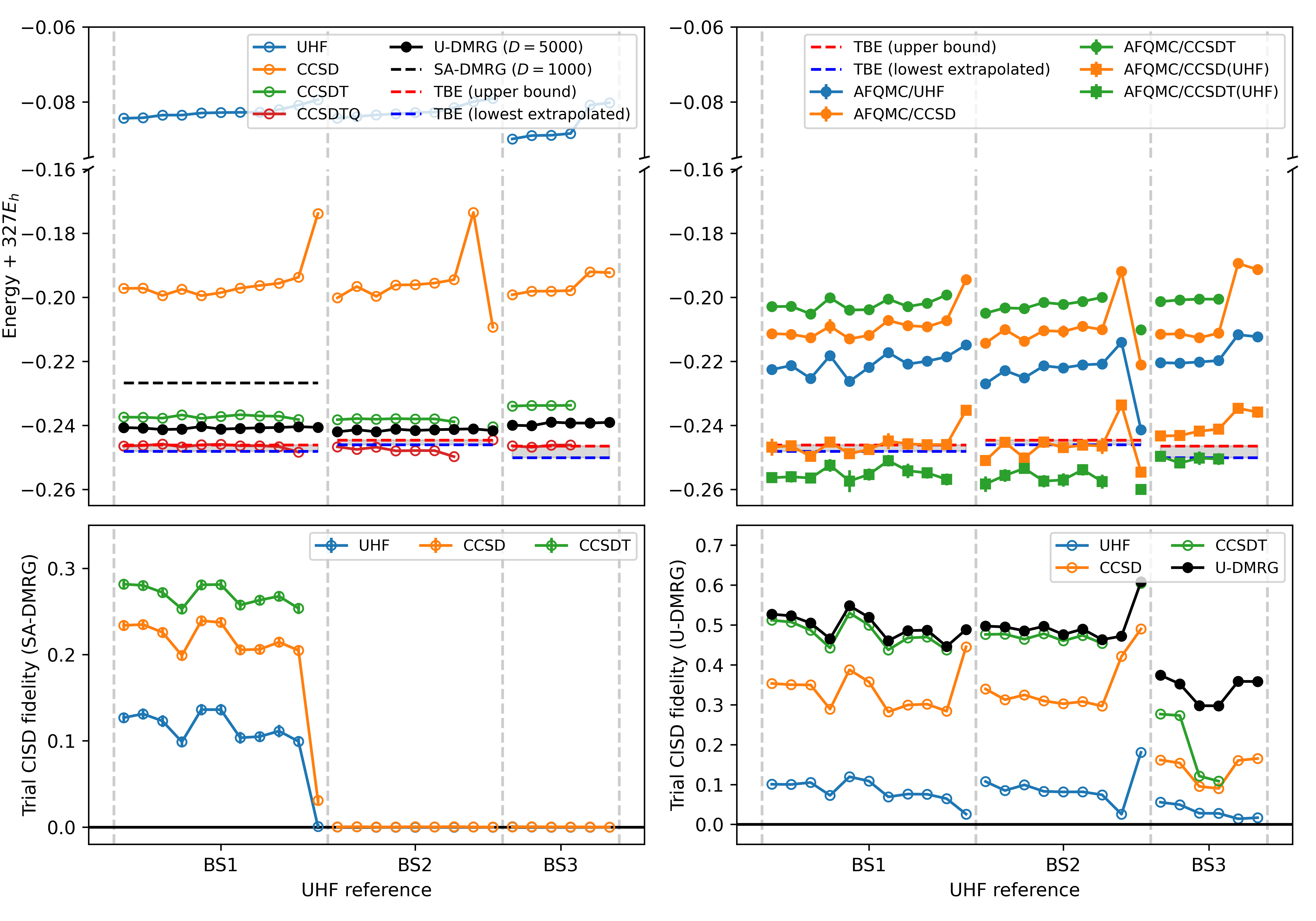}
    \caption{Energies and trial state fidelities for the cubane cluster [4Fe-4S]. The $x$ axes denote individual UHF references, which are grouped into BS1--3 with dividing lines (gray, dashed). Upper: CC trial state and DMRG energies (left) and AFQMC energies (right), where, e.g., CCSD(UHF) denotes a CCSD guiding trial and a UHF measurement trial. Lower: Trial state fidelities in the CISD subspace, computed relative to a SA-DMRG ($D=1000$) reference (left) and U-DMRG ($D=5000)$ references (right). For the U-DMRG references, we also plot their weight in the CISD subspace (black points). Theoretical best estimates (TBEs) for the lowest-lying state in each BS family are from Ref.~\citenum{zhai2026classicalsolutionfemocofactormodel}.}
    \label{fig:fe4}
\end{figure}

Our [4Fe-4S] results are summarized in Figure \ref{fig:fe4}. Starting with the trial energies (upper left), we find that the CC energies converge uniformly as we ascend the hierarchy, like for [2Fe-2S]. In contrast, the AFQMC energies (upper right) exhibit an inverted pattern of convergence, similar to that observed for the dimer. In particular, the phaseless AFQMC energies obtained with a UHF trial are more accurate than those with a CCSD trial, which in turn are more accurate than those with a CCSDT trial. These plots also show theoretical best estimates (TBEs) for each BS family (gray shaded areas) based on CC and DMRG calculations from Ref.~\citenum{zhai2026classicalsolutionfemocofactormodel}. Note that according to these estimates, the BS3 family is the most likely candidate for the true ground state.

As shown in the lower panels, trial fidelities in the CISD subspace are increasing when computed relative to both SA-DMRG (left) and U-DMRG (right) references, mirroring our findings for [2Fe-2S]. The phaseless AFQMC energies are thus again becoming less accurate even as the trial state is improving. The fidelities relative to the SA-DMRG reference (lower left) are non-zero only for BS1 references. This has no significance and only reflects that the DMRG state we used corresponds to a root in the BS1 family. Across all three BS families, we find that fidelities are increasing relative to U-DMRG references (lower right), and that CCSDT trials recover most of the weight in the CISD subspace for BS families 1 and 2. This strongly suggests that the correlated CC trial states generally improve upon the mean-field description.

Although this system is too large to estimate walker fidelities, we can again consider the effect of using the UHF trial for energy measurement. As for [2Fe-2S], this removes the inverted pattern (see upper right) and improves the general accuracy of the phaseless AFQMC energies. However, in this case, the energies appear to overshoot for a CCSDT guiding trial, with the possible exception of the (presumed) ground state family BS3.

\subsection{The FeMo cofactor}
The active space model for the FeMo cofactor comprises 113 electrons in 76 orbitals, including Fe $3d$, Mo $4d$, S $3p$, and C $2s, 2p$, and some ligand orbitals.\cite{li2019electronic} For this system, spectroscopic evidence suggests that the ground state is of overall spin $S = 3/2$ and charge $-1$.\cite{Stappen2020,MUNCK197532} As for the smaller Fe-S clusters, physically meaningful initial UHF densities are obtained by distributing oxidation states and $\alpha$/$\beta$ configurations on each of the Fe and Mo centers. We restrict our attention to a set of 35 filtered UHF references from Ref.~\citenum{zhai2026classicalsolutionfemocofactormodel} obtained by ranking references for each spin-isomer according to energies obtained with high-order CC calculations. These belong to 10 BS families labeled as BS1--10.

\begin{figure}[!htbp]
    \includegraphics[width=\linewidth]{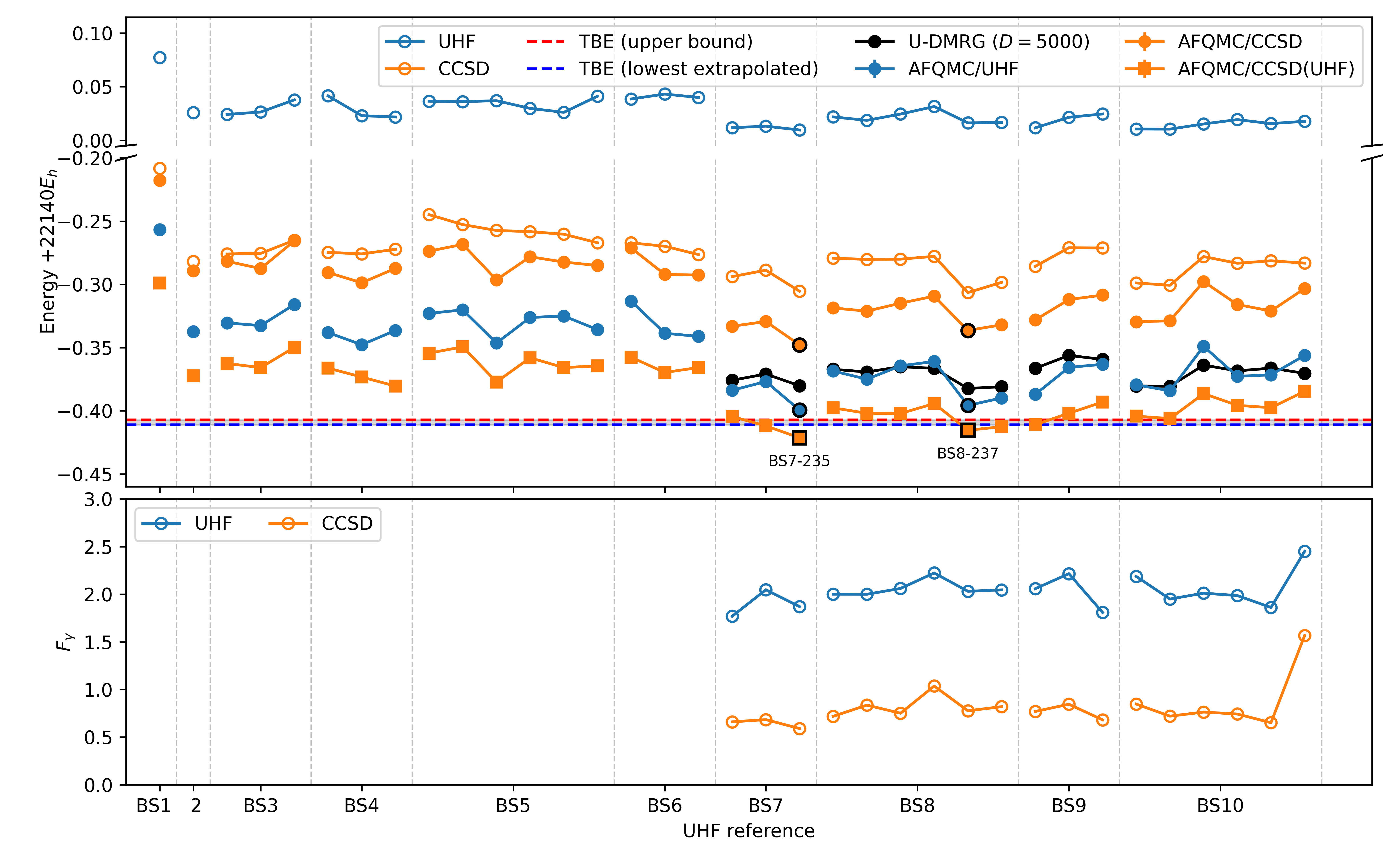}
    \caption{Energies and Frobenius norm-deviations of one-particle density matrices for the FeMo cofactor. This norm is computed as $F_\gamma = \vert\vert \boldsymbol{d}_T - \boldsymbol{d}_0 \vert\vert_F$, where $\boldsymbol{d}_T$ is the spin-summed one-particle density matrix of the trial state and $\boldsymbol{d}_0$ is the corresponding density matrix of U-DMRG ($D=5000$) references obtained for each UHF reference. These DMRG references and the theoretical best estimates (TBEs) are from Ref.~\citenum{zhai2026classicalsolutionfemocofactormodel}.}
    \label{fig:femoco}
\end{figure}

The results for the FeMo cofactor are given in Figure \ref{fig:femoco}. Considering first the trial and AFQMC energies, we see a by now familiar pattern: trial energies are improving when going from UHF to CCSD, whereas the AFQMC energies based on these trials show the opposite trend, with AFQMC/UHF energies being lower in energy and closer to the TBE than the AFQMC/CCSD energies. It is useful to narrow one's attention to the most likely BS families of the true ground state. As shown in Ref.~\citenum{zhai2026classicalsolutionfemocofactormodel}, considerations based on high-level correlated wave function calculations suggest that BS7 and BS8 are the most likely candidate families, with BS7 maximizing the number of anti-ferromagnetic couplings across Fe-S-Fe bonds. Within each family, the most likely candidate references  for the ground state (BS7-235 and BS8-237)\cite{zhai2026classicalsolutionfemocofactormodel} are also denoted in the figure. Across all the BS families, we find that UHF measurement reverses the inverted energy trend, consistent with our findings for [2Fe-2S] and [4Fe-4S]. Focusing in on BS7 and BS8, we find that AFQMC using a CCSD guide and a UHF trial yields the most accurate energies, but still displays a significant spread with errors an order of magnitude higher than that needed to approach chemical accuracy (around 30 mH or 20 kcal/mol).

As for the smaller clusters, one can ask whether the CCSD trial state truly improves upon the UHF trial state. For this system, we do not have a direct fidelity estimate. However, we find that deviations in the one-particle density matrix relative to U-DMRG references ($D=5000$) are reduced for BS7--10 when going from a UHF trial to a CCSD trial. Although not conclusive evidence, this finding and the consistent behavior across the Fe-S clusters suggest that the trial state is becoming more accurate even as the phaseless AFQMC energies (using the standard estimator) become less accurate.

\section{Conclusions} \label{sec:conclusions}
This work investigates the performance of the phaseless AFQMC method in predicting the ground state energy of three active space models of Fe-S clusters ([2Fe-2S], [4Fe-4S], and the FeMo cofactor), applying our recent implementations of high-order CC trial states.\cite{mahajan2025beyond, kjonstad2025}
The use of a hierarchy of CC trial states allows us to systematically probe the accuracy of the method as the quality of the trial state that is used to control the fermion sign problem is improved.
The application to these clusters appeared promising, as
phaseless AFQMC has been argued to perform well on systems containing transition metals, and recent results with CCSD trial states had shown encouraging results.\cite{mahajan2025beyond}

However, our study instead documents some surprising limitations of the method as applied to Fe-S clusters when using broken symmetry trials.
Typically, phaseless AFQMC significantly improves upon the energy of the underlying trial state.\cite{motta2018ab,lee2022twenty}
For the Fe-S clusters studied here, the opposite can occur: the phaseless AFQMC energies can become less accurate than the trial, even as the trial state and its projected energy are significantly improved, and in some cases even as both the trial state and the walker wave function are improved. This behavior is a consequence of the non-variational nature of the method's energy estimator, in which the trial state represents the bra and the walker wave function the ket; improving either or both does not guarantee an improvement in the predicted energy. For all clusters, our results suggest that the trial states improve as we ascend the CC hierarchy, and for [2Fe-2S] this coincides with a simultaneous improvement in the walker wave function; for the larger clusters, the accuracy of the walkers is difficult to assess as it requires extensive sampling. These findings appear moreover to be unrelated to the use of spin-unrestricted trial states and walkers, as the energy trends persist even when one stabilizes the broken-symmetry state in [2Fe-2S] by applying a fictitious local staggered spin-Zeeman field. Overall, our findings should caution against assuming that phaseless AFQMC is suited for systems of this kind
when unrestricted Hartree-Fock and low-order coupled cluster trial wavefunctions are used.
In particular, the often quite accurate ground state energies obtained with a UHF trial (also observed in this work) may be a product of a favorable error cancellation and should not, on their own, be taken as evidence of an accurate representation of the ground state.

Our findings also raise questions regarding the optimal choice of energy estimator.
The estimator probes different excitation sectors in the walker wave function depending on the trial state. This stands in contrast to CC theory, in which the energy expression always probes the components of the CC state in the subspace of single and double excitations relative to the HF reference. In AFQMC, the trial state can similarly project out high-order components in the walkers and may thus suppress errors in high-order excitation sectors.
For example, quadruple excitations in the walkers are ``visible'' to the estimator when using a CISD trial state (e.g., obtained from CCSD), while only up to double excitations are visible for a UHF trial.
Consistent with this, phaseless AFQMC energies improve on the Fe-S clusters when using a UHF measurement trial while guiding the walkers with a CC trial state.
However, our results suggest some caution about the suitability of this approach, since for [4Fe-4S] the results appear overall less accurate with a CCSDT guide compared with a CCSD guide for some of the broken-symmetry families.

Despite all the above limitations, the challenging nature of Fe-S clusters for all established electronic-structure methods should be kept in mind. Compared with other approaches at similar computational scaling, the phaseless AFQMC method provides competitive ground state energies also on these systems. Nevertheless, the observed inverted behavior of phaseless AFQMC---better trial, worse energy---stands in clear contrast to the systematic and uniform convergence of the CC hierarchy on these systems. This seems a significant observation, and these systems may serve as a useful failure case to guide the further development of phaseless AFQMC and to better understand its suitability for similar systems.

\begin{acknowledgement}
We thank Don Danilov, Brad Ganoe, and Ankit Mahajan for helpful discussions and Ankit Mahajan for assistance with AD-AFQMC. EK acknowledges helpful discussions with Yichi Zhang. Work by EK and GKC was supported by the US Department of Energy (DOE), Office of Science, via award no. DE-SC0018140C. The Flatiron Institute is a division of the Simons Foundation. JS acknowledges support from the Robert A. Welch Foundation, Award Number C-2212. SS was supported by the DOE grant DE-SC0025943.
\end{acknowledgement}

\begin{suppinfo}
Additional details about the estimation of walker wave function fidelities, active space compression for [2Fe-2S], occupation strings for all clusters, and the symmetry-breaking spin-Zeeman field for [2Fe-2S]. Reported data (e.g., energies and fidelities) are provided in a separate data repository, along with integral files (\textsc{FCIDUMP}) for the reduced (18e, 14o) active space model for [2Fe-2S].\cite{zenodo}
\end{suppinfo}

%%%REFERENCES%%%
\bibliography{paper}



\section*{Supplementary material}

# S1 Fidelity estimation for the walker wave function

We are concerned with estimating the *fidelity* of the walker wave function $|\psi\rangle$ with respect to a reference state $|\psi_0\rangle$, which we define in this work as

$$f = \frac{|\langle\psi_0|\psi\rangle|}{\langle\psi_0|\psi_0\rangle^{1/2}\langle\psi|\psi\rangle^{1/2}}, \quad |\psi\rangle = \sum_I w_I \frac{|\phi_I\rangle}{\langle\psi_T|\phi_I\rangle}. \tag{1}$$

As the reference state is converted into a configuration interaction (CI) expansion, $|\psi_0\rangle = \sum_I c_I|I\rangle$, its norm is readily computed as $d_0 = \langle\psi_0|\psi_0\rangle^{1/2} = (\sum_I c_I^2)^{1/2}$. We can therefore focus on the estimation of the terms in the numerator and denominator involving the walkers, i.e.

$$n = \langle\psi_0|\psi\rangle = \Big|\sum_I w_I \frac{\langle\psi_0|\phi_I\rangle}{\langle\psi_T|\phi_I\rangle}\Big| \tag{2}$$

and

$$d = \langle\psi|\psi\rangle = \sum_{IJ} \frac{w_I w_J}{\langle\phi_I|\psi_T\rangle\langle\psi_T|\phi_I\rangle}\langle\phi_I|\phi_J\rangle. \tag{3}$$

Note that $I$ correspondings to the compound index $(n, k)$, where $n$ denotes the timestep and $k$ denotes the walker. In our evaluations, $n$ denotes a specific block as we store walkers at the end of a block containing a given number of timesteps (in our case, 50). As is usual practice, we remove the initial equilibration phase from $|\psi\rangle$ as we are interested in the steady-state distribution of the walker ensemble, and not the initial dynamics.

Since exact evaluation of these quantities is computationally expensive, we adopt an importance sampling approach to estimate $n$ and $d$ more efficiently. In particular, we estimate the sum in $n$ by sampling indices $\{I_k\}$ from the distribution

$$p_I = \frac{w_I}{|\langle\psi_T|\phi_I\rangle|}\Big/\sum_J \frac{w_J}{|\langle\psi_T|\phi_J\rangle|}, \tag{4}$$

and to evaluate $d$, we sample indices $\{I_k, J_k\}$ independently from the same distribution, i.e. we effectively sample from the product distribution $p_{IJ} = p_I p_J$. Note that $\{p_I\}$ is readily evaluated, requiring only the overlaps $\langle \psi_T | \phi_J \rangle$ between the trial state and the walkers. As with other local quantities like the local energy, we perform real-axis projection for each $I$ instead of carrying the phase and taking the modulus after summation over $I$. In tests on the reduced [2Fe-2S] (18e, 14o) model, using UHF trials, we find that the overall complex phase is zero to within stochastic error, and that the local real-axis projection has a negligible effect on the estimated fidelity.

At this point, we should distinguish two sources of error in the $f$ estimate: (a) error from the finite number of walker samples and (b) error from the finite number of importance samples used to evaluate $n$ and $d$ for a *given* sample of walkers. Let us consider (b) first. In practice, we use a sufficiently high number of MC samples to ensure the error due to importance sampling is small compared to the uncertainty in $f$ due to the finite number of walkers. Since we draw samples from $p$ independently for $n$ and $d$, we assume that the covariance term can be neglected, so that the resulting delta-error estimate in $f_W$, conditioned on a sample of walkers $W = \{I\}$, is

$$\sigma^2_{f_W} = \frac{1}{d_0^2}\Big(\frac{\sigma_n^2}{d} + \frac{n^2}{4d^3}\sigma_d^2\Big). \tag{5}$$

For the inherent uncertainty (a) due to the finite set of walkers, we estimate the uncertainty by partitioning the set of walkers into $K = 10$ independent walker windows $W_i$ (blocks 1–5000, 5001–10000, etc.) and estimating the variation across these windows. That is, we compute the mean estimated fidelity across the windows,

$$\mu_{f_W} = \frac{1}{K}\sum_{i=1}^{K} f_{W_i}, \tag{6}$$

from which we obtain an estimated variance across the windows, as

$$\sigma^2_{f_W} = \frac{1}{K-1}\sum_{i=1}^{K}(f_{W_i} - \mu_{f_W})^2. \tag{7}$$

As we are specifically interested in the between-window variance, we subtract the average inner-window variance due to importance sampling, yielding the inter-window variance estimate as

$$\tilde{\sigma}^2_{f_W} = \sigma^2_{f_W} - \frac{1}{K}\sum_{i=1}^{K}\sigma^2_{f_{W_i}}. \tag{8}$$

(Note that this may become negative if inner-window noise dominates, but we ensure a sufficient number of samples to avoid this.) Denoting the full sample of walkers as $S = \{I\}$, where $S = \cup_{i=1}^{K} W_i$, we then find that an estimated variance by summing the two sources of error,

$$\sigma^2_f = \sigma^2_{f_S} + \frac{\tilde{\sigma}^2_{f_W}}{K}. \tag{9}$$

Note that we divide by $K$ because $\tilde{\sigma}^2_{f_W}$ estimates the variance of a single window's fidelity, whereas we require the variance of the mean over $K$ independent windows. Estimated errors in $f$ are then reported as $\sigma_f$.

This procedure requires a sufficient number of walkers in each window, as near-convergence of $f$ is needed to avoid significant bias in the smaller window-samples $W_i$ compared with the full set of walkers $S$. This motivates the choice of $K = 10$ windows, as we find that for the [2Fe-2S] (18e, 14o) model that convergence of $f$ is reached after about 5000 blocks (each containing 200 walkers) and we have sampled $10 \times 5000 = 50\,000$ blocks. To confirm the cumulative convergence of $f$, we have estimated the fidelity for an increasing number of blocks. The results are given in Figure 1. Note that for all calculations, we have used $M_n = 10^5$ MC samples for the evaluation of $n$ and $M_d = 8 \times 10^7$ MC samples for the evaluation of $d$.

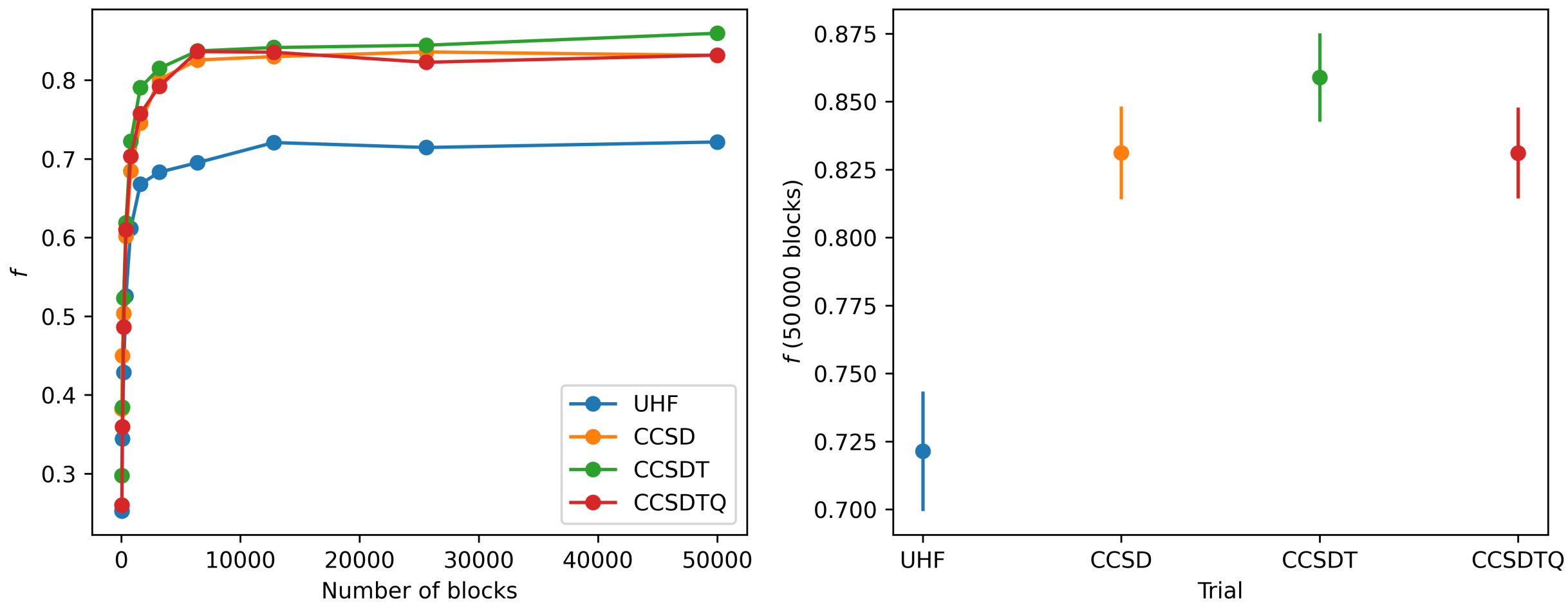


Figure 1: Cumulative convergence of the fidelity $f$ (left) and total fidelity estimate (right) for [2Fe-2S] (18e, 14o).

The walker ensemble's higher fidelity relative to the broken-symmetry trial states likely arises because the spin-unrestricted walkers become restricted ($S = 0$) during imaginary-time propagation and represent a singlet in the long-time limit (see Figure 2, upper). Note that the spin-purity of the walkers does not have a significant effect on the local energies (see Figure 2, lower).

# S2 Reduced active space models for the [2Fe-2S] cluster

Based on the (30e, 20o) active space model for [2Fe-2S] by Li and Chan,[1] we construct compressed active spaces that preserve the main qualitative features of the parent model. To perform this compression, we make use of the ground state density matrix for the (30e, 20o) model (computed with $D = 8000$ spin-adapted DMRG) and apply a occupation cutoff $n_o$ to remove the natural orbitals that can be considered approximately frozen (i.e., has an occupation close to 2). For a given cutoff, we transform the Hamiltonian integrals into the truncated orbital basis and compute the constant frozen core energy contributions. We find two reduced active spaces that preserve the main features of the full model, $n_o < 1.98$ (22e,

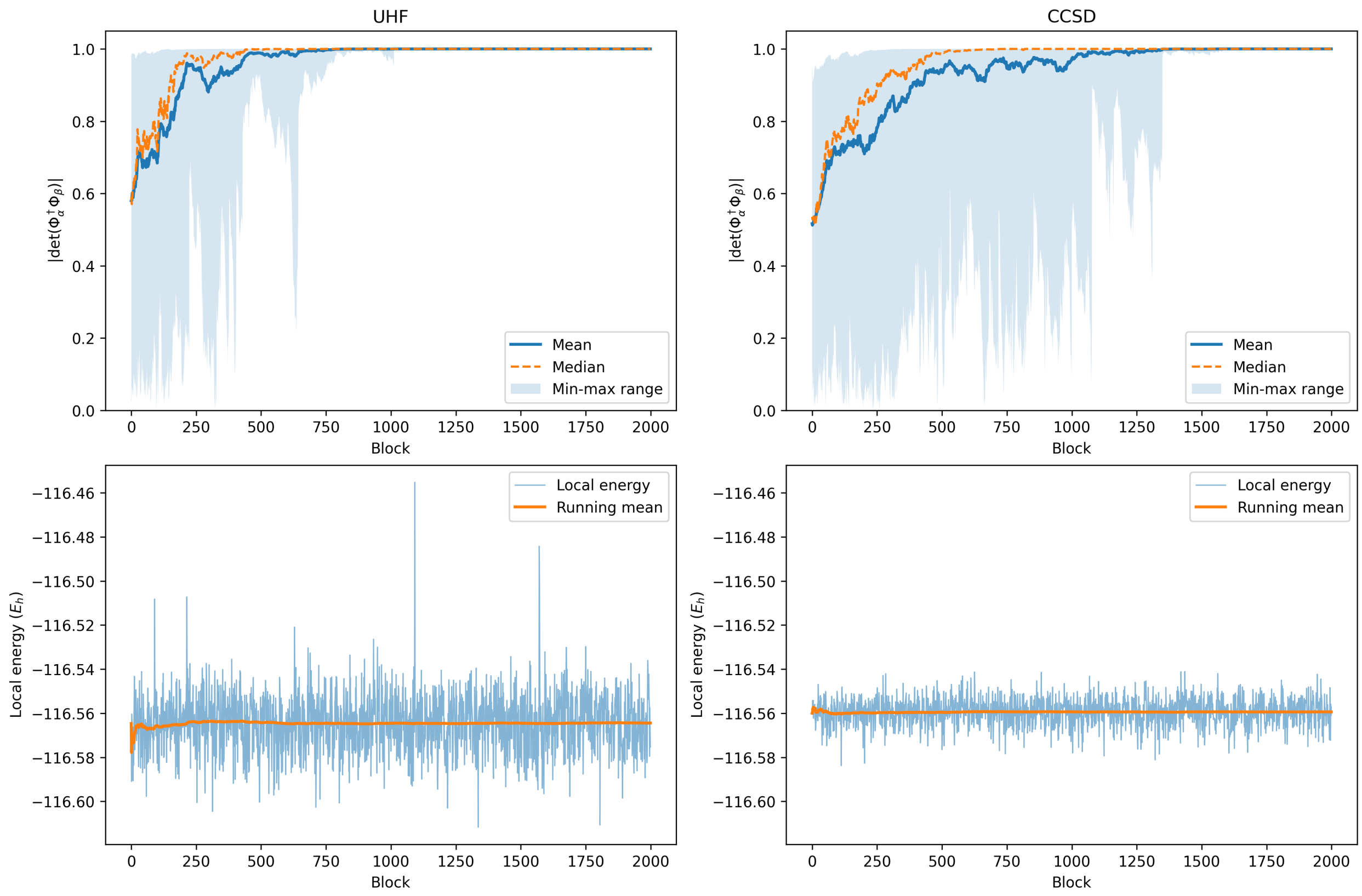


Figure 2: $\alpha/\beta$ walker overlaps (top) and associated local energies (bottom) for [2Fe-2S] (18e, 14o) using UHF (left) and CCSD (right) trial states.

16o) and $n_o < 1.97$ (18e, 14o). In this work, we use the (18e, 14o) model to estimate walker wave function fidelities for the iron dimer.

Our results are summarized in Figure 3, where we show energies obtained using coupled cluster trial states for different cutoff values and active space sizes. We find that the trends are almost exactly preserved for the (22e, 16o) and (18e, 14o) active spaces, when compared with the full (30e,20) model, while significant differences are observed when further truncation is made (14e, 12o). For future reference and reproducibility, we provide FCIDUMP files for the (18e, 14o) and (22e, 16o) models of the [2Fe-2S] dimer, along with initial densities used to converge the UHF solutions, in the Supplementary data repository.[2]

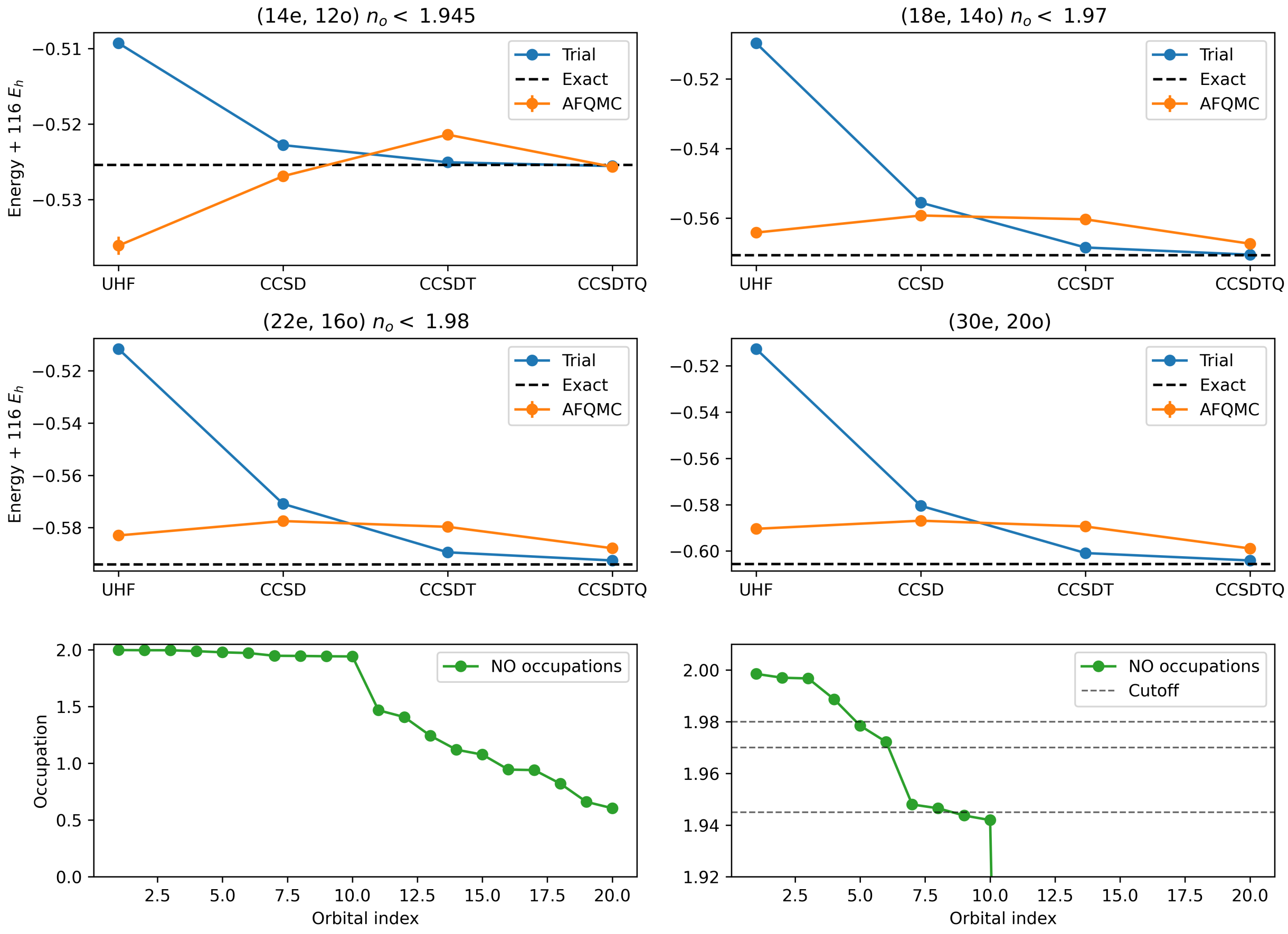


Figure 3: Upper and middle panels: Trial dependence of the phaseless AFQMC energies for various reduced active space models for [2Fe-2S]. Lower panels: Natural orbital occupations for spin-adapted DMRG density matrix (left), including a zoomed-in region including cutoff values (gray dashed lines).

# S3 Initial spin-occupation strings for convergence of mean-field references

For each cluster, we converge UHF by starting from physically motivated density matrices that reflect local atomic spins, Fe (and Mo) oxidation states, and the overall total spin of each system ($S = 0$ in the case of [2Fe-2S] and [4Fe-4S], and $S = 3/2$ in the case of the FeMo cofactor). These densities are used as initial guesses in the second-order Newton solver in PySCF.[3] In the case of [2Fe-2S] with 2Fe(III), each of the iron atoms is assigned a $d^5$ configuration with parallel spins. This gives the unique initial occupation string "22aaaaa222222bbbbb22" up

to symmetry, where the bridging orbitals are doubly occupied ("2") and "a" and "b" refers to $\alpha$ and $\beta$ occupation, respectively. For [4Fe-4S] with 2Fe(III)-2Fe(II), there are many ways to distribute the oxidation states among the four Fe atoms as well as to select which $d$ orbital is doubly occupied for the two Fe atoms with oxidation state II. Considering all possible permutations, we have converged 26 unique broken-symmetry solutions. For both [2Fe-2S] and [4Fe-4S], we list the initial spin-density configurations in Table 1.

Table 1: Initial spin-occupation strings for the [2Fe–2S] and [4Fe–4S] clusters.

| Cluster | ID | Initial occupation string |
|---|---|---|
| 2Fe-2S | 0 | 22aaaaa222222bbbbb22 |
| 4Fe-4S (BS1) | 0 | 222aaaaaaaaa222222222222bbbb2bbbbb22 |
| | 1 | 222aaaaaaaaa2222222222222bbbbbbbbb22 |
| | 2 | 222aaaaaaaaa222222222222b2bbbbbbbb22 |
| | 3 | 222aaaaaaaaa222222222222bbbbbb2bb22 |
| | 4 | 22aaaaa2aaaa222222222222b2bbbbbbbb22 |
| | 5 | 22a2aaaaaaaa222222222222bb2bbbbbbb22 |
| | 6 | 22aaaaa2aaaa2222222222222bbbbbbbbb22 |
| | 7 | 22a2aaaaaaaa222222222222b2bbbbbbbb22 |
| | 8 | 22a2aaaaaaaa2222222222222bbbbbbbbb22 |
| | 9 | 22aaaaaaaaaa222222222222bbbbbb2bb22 |
| | 10 | 22aaaa2aaaaa2222222222222bbbbbbbbb22 |
| 4Fe-4S (BS2) | 11 | 222aaaab2bbb222222222222aaaaabbbbb22 |
| | 12 | 222aaaa2bbbb222222222222aaaaabbbbb22 |
| | 13 | 22a2aaabb2bb222222222222aaaaabbbbb22 |
| | 14 | 22aaa2abbb2b222222222222aaaaabbbbb22 |
| | 15 | 22a2aaab2bbb222222222222aaaaabbbbb22 |
| | 16 | 22a2aaa2bbbb222222222222aaaaabbbbb22 |
| | 17 | 22aaaaabbbbb222222222222aaaa22bbbb22 |
| | 18 | 22a2aaabbbb2222222222222aaaaabbbbb22 |
| | 19 | 22aaaaabbb2b222222222222a2aaabbbbb22 |
| 4Fe-4S (BS3) | 20 | 222aaaab2bbb222222222222bbbbbaaaaa22 |
| | 21 | 22a2aaa2bbbb222222222222bbbbbaaaaa22 |
| | 22 | 222aaaa2bbbb222222222222bbbbbaaaaa22 |
| | 23 | 22a2aaab2bbb222222222222bbbbbaaaaa22 |
| | 24 | 22aa2aabb2bb222222222222bbbbbaaaaa22 |
| | 25 | 22aaaaabbbbb222222222222bb2bbaa2aa22 |

For the FeMo cofactor model, guesses are generated similarly from physically motivated

initial densities, with a further filtering procedure based on coupled cluster energetics. This filtering procedure is described in Zhai et al.[4] The initial spin-density configurations are listed in Table 2 for ease of reference.

Table 2: Initial occupation strings used to converge broken-symmetry UHF for FeMo-co.[4]

| BS | ID | ID[4] | Initial occupation string |
|---|---|---|---|
| 1 | 0 | 34 | 22aaaaa222222222aa2aa2aaaaaaaaa2222222222222bb2bbbbbbbbbbb222222b222ba00222 |
| 2 | 1 | 15 | 222aaaa222222222bbb2bbbbbbbbbb22222222222222aaaaaaaaaaaaaa222222b222ba00222 |
| 3 | 2 | 0 | 22bbbbb222222222b2bbbbbbbbaaa2a2222222222222aaaaaaaaaaaa2aa222222b222ba00222 |
| 3 | 3 | 1 | 22bbb2b222222222bbbbbaaaaabbbbb2222222222222aaa2aaaaaaaa2aa222222b222ba00222 |
| 3 | 4 | 5 | 22bbbb2222222222a2aaabbbbbbbbbb22222222222222aaaaaaaaaaaaaa222222b222ba00222 |
| 4 | 5 | 24 | 222aaaa222222222b2bbbaa2aaaaaaa2222222222222aaaaabbbbbbbbbb222222b222ba00222 |
| 4 | 6 | 29 | 22aaaa2222222222aaaaa2bbbbaaaaa2222222222222bbbbb2aaaabbbbb222222b222ba00222 |
| 4 | 7 | 31 | 22a2aaa222222222aaa2aaaaaabb2bb2222222222222bbbbbbbbbbaaaaa222222b222ba00222 |
| 5 | 8 | 22 | 22aaaaa2222222222bbbbaa2aaaaaaa2222222222222bbbbbbbbbbaaa2a222222b222ba00222 |
| 5 | 9 | 23 | 222aaaa222222222b2bbbaaaaaaaaaa2222222222222bbbbb2aaaabbbbb222222b222ba00222 |
| 5 | 10 | 28 | 22aaa2a222222222aaaaabbbb2a2aaa2222222222222bbbbbbbbbbaaaaa222222b222ba00222 |
| 5 | 11 | 30 | 22aaaaa222222222aa2aabbbbbaaaa22222222222222aaaaa2bbbbbbbbb222222b222ba00222 |
| 5 | 12 | 32 | 22aaaaa222222222a2aaaaaaaab2bbb2222222222222bbbbb2aaaabbbbb222222b222ba00222 |
| 5 | 13 | 33 | 222aaaa222222222a2aaaaaaaab2bbb2222222222222aaaaabbbbbbbbbb222222b222ba00222 |
| 6 | 14 | 12 | 22bbbbb222222222aa2aaaaaaa2aaaa2222222222222b2bbbbbbbbaaaaa222222b222ba00222 |
| 6 | 15 | 13 | 22bbb2b222222222aaaaaaaaaaaa2aa2222222222222bbbbb2aaaabbbbb222222b222ba00222 |
| 6 | 16 | 14 | 22bbbb2222222222aaa2aaaaaaaa2aa2222222222222aaaaabbbbbbbbbb222222b222ba00222 |
| 7 | 17 | 18 | 222aaaa222222222bb2bbbbbbbaaaaa2222222222222aa2aaaaaaabbbbb222222b222ba00222 |
| 7 | 18 | 20 | 22aaaa2222222222bbbbaaaaabbbbb2222222222222aaaaabbbbbaaaa2222222b222ba00222 |
| 7 | 19 | 25 | 22a2aaa222222222aaaaabbb2bbbbbb2222222222222bbbbbaaaaaaa2aa222222b222ba00222 |
| 8 | 20 | 16 | 22a2aaa222222222bbb2bbbbbbaaaaa2222222222222bbbbbaaaa2aaaaa222222b222ba00222 |
| 8 | 21 | 17 | 222aaaa222222222bbb2bbbbbbaaaaa2222222222222aaaaabbbbb2aaaa222222b222ba00222 |
| 8 | 22 | 19 | 22aaa2a222222222bbb2baaaaabbbbb2222222222222bbbbbaaaaaa2aaa222222b222ba00222 |
| 8 | 23 | 21 | 22aaaa2222222222b2bbbaaaaabbbbb2222222222222aa2aaaaaaabbbbb222222b222ba00222 |
| 8 | 24 | 26 | 22aa2aa222222222aaaaabbbbbbbbbb2222222222222a2aaabbb2baaaaa222222b222ba00222 |
| 8 | 25 | 27 | 22aaaa2222222222aaaaabbb2bbbbbb2222222222222aaaaaaaaa2bbbbb222222b222ba00222 |
| 9 | 26 | 2 | 22bbbbb222222222bbbbba2aaa2aaaa2222222222222bb2bbaaaaaaaaaa222222b222ba00222 |
| 9 | 27 | 7 | 22bbbbb222222222aaa2abbbbbaaaaa2222222222222aaa2a2bbbbaaaaa222222b222ba00222 |
| 9 | 28 | 11 | 22bbb2b222222222a2aaaaaaaabbbbb2222222222222aaaa2aaaaabbbbb222222b222ba00222 |
| 10 | 29 | 3 | 22bb2bb222222222bbbbbaaaaa2aaaa2222222222222aaaa2bbbbbaaaaa222222b222ba00222 |
| 10 | 30 | 4 | 22b2bbb222222222bbbbb2aaaaaa2aa2222222222222aaaaaaaaaabbbbb222222b222ba00222 |
| 10 | 31 | 6 | 22bbbbb222222222aaa2a2bbbbaaaaa2222222222222bbbbbaaaaa2aaaa222222b222ba00222 |
| 10 | 32 | 8 | 22bbbbb222222222a2aaab2bbbaaaaa2222222222222aaaaaa2aaabbbbb222222b222ba00222 |
| 10 | 33 | 9 | 22bbbbb2222222222aaaaaaaaabbbbb22222222222222bbbbaaaaaaaaa2a222222b222ba00222 |
| 10 | 34 | 10 | 22bbbbb222222222aaaaa2aaaabb2bb22222222222222aaaabbbbbaaaaa222222b222ba00222 |

For the specific orbital ordering in the three iron-sulfur clusters, and a complete list of the included active-space orbitals, we refer the reader to the original papers.[1,5]

## S4 Symmetry-breaking field for the [2Fe-2S] cluster

To assess the impact of using spin-unrestricted trial states and walkers, we apply a fictitious spin-Zeeman term with a staggered pattern on the $d$ orbitals localized on the two Fe centers in [2Fe-2S]. The effect of the field is to break the $S^2$ symmetry of the ground state while preserving $M_S$, and to stabilize the symmetry-broken mean-field solution. For a sufficiently strong field, the ground state of the modified Hamiltonian acquires broken-symmetry character, and the trial states become more accurate relative to the reference (higher fidelity). More importantly for our purposes, the field allows us to probe whether the observed energy trend in [2Fe-2S] is related to the loss of $S^2$ symmetry, and in particular whether it is related to the use of spin-unrestricted trials and walkers.

Denoting the two Fe centers as $\mathrm{Fe}_A$ and $\mathrm{Fe}_B$, we apply the following field term to the one-body Hamiltonian,

$$\begin{aligned} h^{\alpha}_{pp} &= -\lambda(\delta_{p\in A} - \delta_{p\in B}) \\ h^{\beta}_{pp} &= +\lambda(\delta_{p\in A} - \delta_{p\in B}), \end{aligned} \tag{10}$$

where $\lambda$ is the field-strength and $p \in X$ denotes that $p$ is a $d$-orbital located on the $\mathrm{Fe}_X$ center. In effect, the field shifts the $\alpha$ and $\beta$ orbital energies in opposite directions on each Fe center, with the pattern reversed between $\mathrm{Fe}_A$ and $\mathrm{Fe}_B$, thereby stabilizing the Fe↑–Fe↓ or Fe↓–Fe↑ broken-symmetry solution (depending on the sign of $\lambda$).

To apply the symmetry-breaking field to the reduced (18e, 14o) model of [2Fe-2S], we first express the one-body Hamiltonian $h$ in Eq. (10) in the original (30e, 20o) active orbital space, and then transform the $h$ contribution into the reduced orbital basis of the (18e, 14o) active space model.

Results using the standard AFQMC energy estimator are given in the main text (see Figure 5 in the manuscript). Below, we include energies obtained using a CCSD guide and a UHF measurement, denoted AFQMC/CCSD(UHF). The results show that the accuracy of

the approach is overall higher than the standard AFQMC/CCSD energies for the considered field strengths $\lambda$.

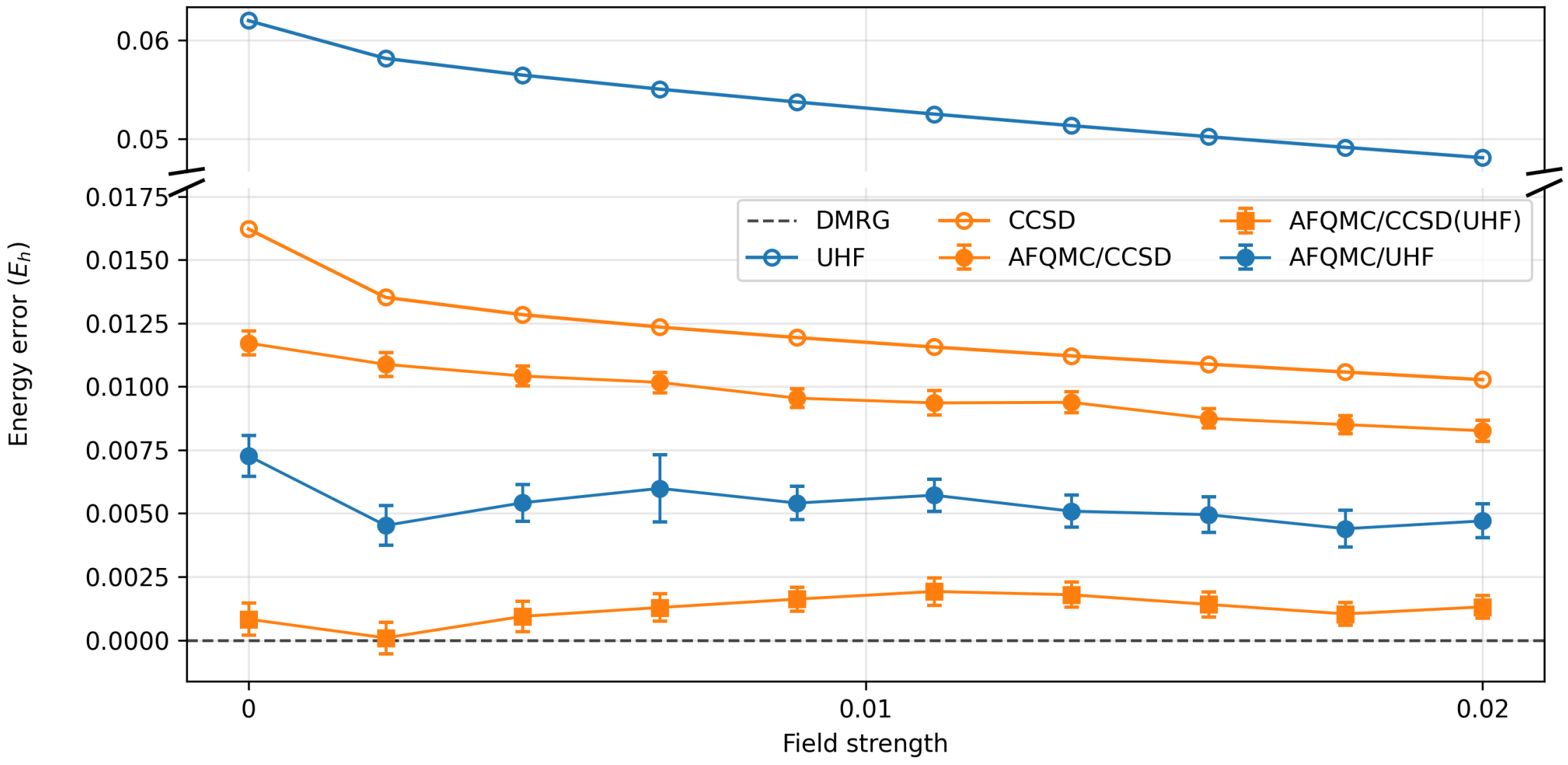


Figure 4: Energies for [2Fe-2S] (18e, 14o) with a symmetry-breaking spin-Zeeman field that stabilizes the $Fe_A(\uparrow)$–$Fe_B(\downarrow)$ broken-symmetry solution. Field strengths are in a.u. Errors are computed relative to DMRG ($D = 3000$) references.

## S5 Local energy variance for the [2Fe-2S] cluster with mean-field measurement

In this work, we consider the application of a UHF measurement trial and a CC guiding trial. In such an approach, one concern is increased variance in the energy estimator. Figure 5 shows local energies and local energy variances for three AFQMC calculations on the [2Fe-2S] (18e, 14o) system: AFQMC/UHF, AFQMC/CCSD, and AFQMC/CCSD(UHF). Note that for this system, the variance of AFQMC/CCSD(UHF) is higher than AFQMC/CCSD but lower than AFQMC/UHF.

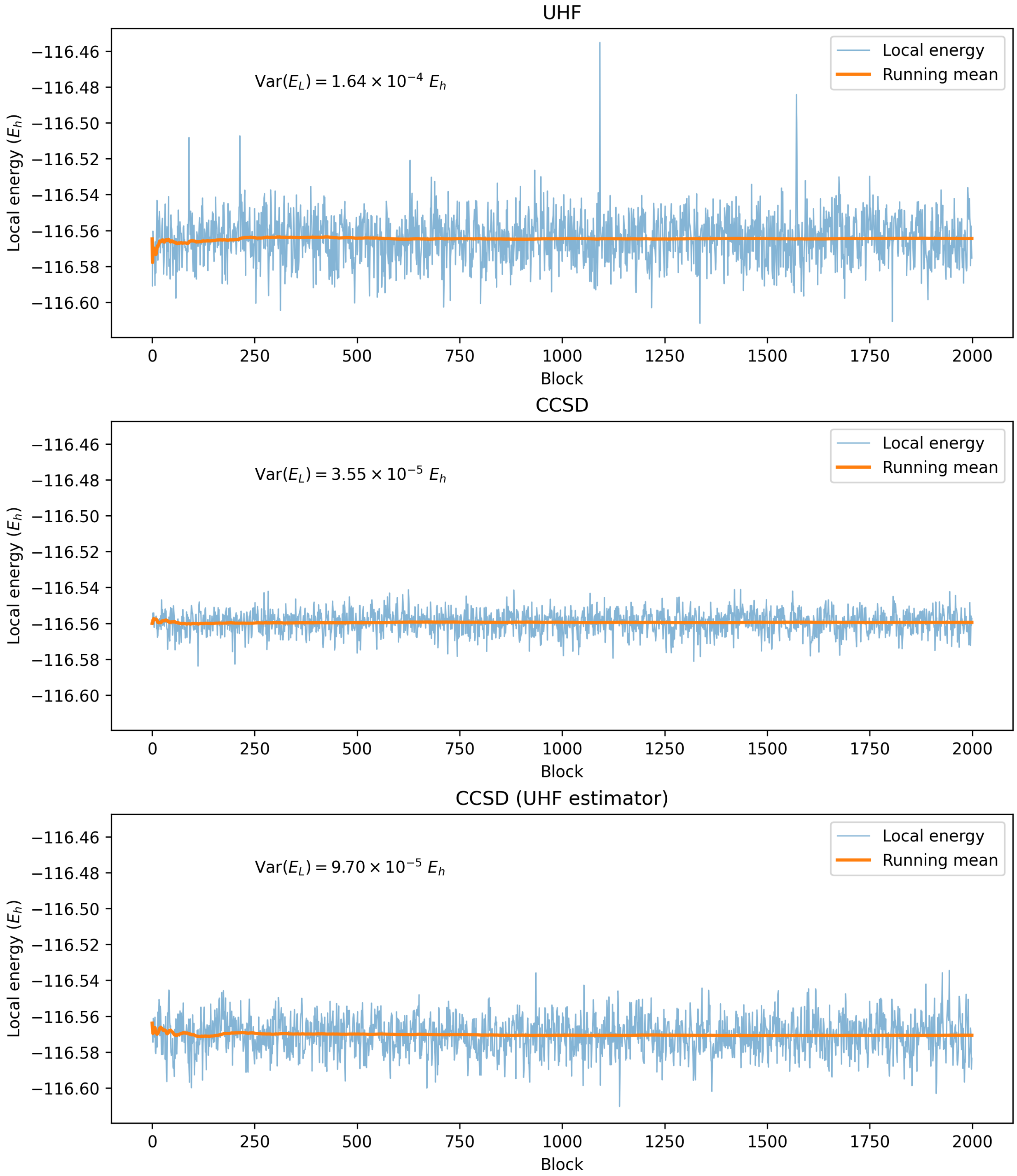


Figure 5: Local energies and local energy variance for [2Fe-2S] (18e, 14o) with AFQMC/UHF, AFQMC/CCSD, and AFQMC/CCSD(UHF).



\end{document}